\documentclass[12pt,letterpaper]{article}
\pdfoutput=1
\usepackage{setspace}
\usepackage{color}
\usepackage{amsmath,mathtools,amssymb,slashed,url}
\usepackage{amsthm}
\usepackage{graphicx,epsfig}
\usepackage{array}
\usepackage{subfigure}
\usepackage{pifont}
\usepackage{empheq}
\usepackage{cancel}
\usepackage{amsfonts}
\numberwithin{equation}{section}
\usepackage{bm}
\usepackage{xcolor}
\usepackage[linktoc=all,breaklinks=true,colorlinks=true,linkcolor=purple,urlcolor=purple,citecolor=blue,%
  pdftitle={Partition Functions of Chern-Simons Theory on Handlebodies by Radial Quantization},
  pdfsubject={hep-th},%
  pdfauthor={Massimo Porrati, Cedric Yu}]{hyperref}
\usepackage{dsfont}
\usepackage{pbox}
\usepackage{float}
\usepackage[nottoc,notlot,notlof]{tocbibind}

\numberwithin{equation}{section}

\newtheorem*{theorem}{Theorem}
\def\b{\beta}

\def\Tr{\mbox{Tr}\,}

\newcommand{\fr}[1]{\mathfrak{#1}}

\def\beq{\begin{equation}}
\newcommand{\eeq}[1]{\label{#1}\end{equation}}
\def\bea{\begin{eqnarray}}
\newcommand{\eea}[1]{\label{#1}\end{eqnarray}}
\def\ba{\begin{array}}
\def\ea{\end{array}}
\renewcommand{\Im}{\mbox{Im}\,}

\date{}
\begin{document}
\def\draftnote#1{{\color{red}#1}}

\pagenumbering{roman}
\begin{titlepage}
\vspace{20pt}
\begin{center}
{\huge Partition Functions of Chern-Simons Theory on Handlebodies by Radial Quantization}

\vspace{18pt}

{\large Massimo Porrati$^\natural$~\footnote{E-mail: \href{mailto:massimo.porrati@nyu.edu}{massimo.porrati@nyu.edu}} and Cedric Yu$^\natural$~\footnote{E-mail: \href{mailto:cedric.yu@nyu.edu}{cedric.yu@nyu.edu}}}

\vspace{12pt}

{$^\natural$ \em Center for Cosmology and Particle Physics, \\ Department of Physics, New York University, \\726 Broadway, New York, NY 10003, USA}

\vspace{12pt}

\end{center}

\abstract{We use radial quantization to compute Chern-Simons partition functions on handlebodies of arbitrary genus. 
The partition function is given by a particular transition amplitude between two states which are
defined on the Riemann surfaces that define the 
(singular) foliation of the handlebody. The final state is a coherent state while on the initial state the holonomy 
operator has zero eigenvalue. 
The latter choice encodes the constraint that the gauge fields must be regular everywhere inside the handlebody. By requiring that the 
only singularities of the gauge field inside the handlebody must be compatible with Wilson loop insertions, we find that the Wilson  loop
 shifts the holonomy of the initial state. Together with an appropriate choice of normalization, this procedure selects a unique state in the Hilbert space obtained from a K\"ahler quantization of the 
theory on the constant-radius Riemann surfaces. Radial quantization allows us to find the partition functions of Abelian Chern-Simons 
theories
 for handlebodies of arbitrary genus. For non-Abelian compact gauge groups, we show that
 our method reproduces the known partition function at genus one.}
\newpage
\end{titlepage}

\pagenumbering{arabic}


\section{Introduction}\label{intro}
Chern-Simons gauge theory connects many different topics in mathematics and physics. On closed manifolds
it is a topological theory that can be used to compute knot invariants~\cite{Witten:1988hf}, while on manifolds with boundaries it
acquires additional boundary degrees of freedom that connect it to gravity in three dimensions~\cite{Achucarro:1987vz,Witten:1988hc,Deser:1983nh,Deser:1983tn} and to the
theory of the fractional quantum Hall effect~\cite{Susskind:2001fb,Zhang:1992eu}. As remarked in~\cite{Beasley:2005vf}, one intriguing feature distinguishes 
Chern-Simons theory from conventional topological field theories, such as topological Yang-Mills theories on Riemann surfaces or 
four-manifolds: the latter can be interpreted in terms of the cohomology ring of some classical moduli space of connections, while 
Chern-Simons, in general, cannot. In fact Chern-Simons theory is intrinsically a quantum theory that is best described by a
Hilbert space. When the three-manifold on which the theory is defined has special characteristics, the theory simplifies and may 
become computable. One remarkable example is the case of Seifert manifolds studied in~\cite{Beasley:2005vf}. Another case which could lead to
exact computations is that of handlebodies~\cite{Rolfsen}. The latter is  
interesting for various reasons. One of the most fascinating is that in order to test any conjectured holographic dualities relating
 pure gravity in three 
dimensions to a conformal field theory~\cite{Witten:2007kt} (or an ensemble average thereof~\cite{Afkhami-Jeddi:2020ezh,Maloney:2020nni,Cotler:2020ugk}), one would need to 
know the partition function of 
$SL(2,\mathbb{C})$ Chern-Simons theory on a negatively curved manifold, whose boundary is a Riemann surface. Handlebodies are
the simplest such manifolds for a fixed genus of the boundary~\cite{Maloney:2007ud}. 

The reason why one may think that a Chern-Simons theory may be exactly 
soluble on handlebodies is that these spaces are {\em almost}
factorized as the topological product $[0,R]\times \Sigma$. We say ``almost'' because the closed
Riemann surface $\Sigma$ defining the
foliation of the space becomes singular at one of the extrema of the interval
$[0,R]$. The simplest example of this foliation is the
  solid torus handlebody, that is the direct product of a disk $D^2$ and a
  circle $S^1$. Its singular foliation is
  $D^2 \times S^1 \approx [0,R]\times T^2$. The $\approx$ sign means that the
two-torus leaf $T^2=S^1\times S^1$ becomes singular at the end $r=0$ of the interval $[0,R]$,
where one of the two $S^1$ cycles degenerates. By interpreting $r\in[0,R]$ as time, we can quantize the theory and define a 
Hamiltonian that evolves in $r$. This allows us to rewrite the partition function of the theory
as a transition amplitude between some initial state $|i\rangle$ at $r=0$ and some final state $|f\rangle$ at $r=R$. We will
show in this paper that the condition that 
the initial state is a ``shrunken,'' degenerate surface imposes a restriction on the initial state that, combined with the constraints
descending from gauge invariance and the independence of the scalar product from the complex structure, 
completely fixes the partition function.

Let us describe now more precisely the procedure that we shall follow and the organization of this paper. We 
study partition functions of Chern-Simons theory of compact gauge groups on handlebodies using a radial quantization. First, we 
establish the equivalence between three quantities: Euclidean path integrals with holomorphic boundary condition, transition 
amplitudes under radial evolution with a coherent state as the final state, and wave functions integrated over the gauge orbit. Second, 
we map a Wilson loop inserted in a path integral to a ``blown-up" operator defined on the Riemann surface, which in the radial 
quantization acts on a seed wave function and defines an initial state of definite holonomy along the contractible cycles. 
 Together with an appropriate choice of normalization, this procedure singles out a unique vector in the Hilbert space 
obtained by a canonical quantization of Chern-Simons theory on the Riemann surface. Moreover, we find that requiring that such 
``blown-up" operator must be gauge-invariant corresponds to selecting a particular class of framings of the original Wilson loop.
 We are thus able to establish a precise state-operator correspondence associating each vector in the Hilbert space of the canonically quantized Chern-Simons theory on $\Sigma$ to an explicitly computed partition function with insertions of Wilson loops.
  We first consider the Abelian $U(1)$ gauge group on the solid torus, then on handlebodies of arbitrary genus and finally 
  we study general compact simple groups on the solid torus.

In \hyperref[sectionabelian]{Section \ref*{sectionabelian}}, we study the $U(1)$ Chern-Simons theory, first on a 
torus handlebody and then on handlebodies defined by higher-genus Riemann surfaces. In \hyperref[secnonabel]{Section 
\ref*{secnonabel}}, we move on to consider the case of a general
 non-Abelian simple compact Lie group on the torus handlebody. Appendices \hyperref[apptheta]{\ref*{apptheta}} and 
 \hyperref[secquaddiff]{\ref*{secquaddiff}} respectively summarize essential facts about the Riemann 
 theta function and quadratic differentials on a Riemann surface.

\section{The Abelian Case}\label{sectionabelian}
To study Chern-Simons theory with gauge group $U(1)$ on the genus-$g$ handlebody $M$, we define a singular foliation on $M$ as 
$M =\Sigma\times [0,R]$. The constant-radius leaves are closed Riemann surfaces, $\Sigma$, and the initial surface $\Sigma_0$ at 
$r=0$ is degenerate. The final surface $\Sigma_R$ is at $r=R$. On $\Sigma$ we specify the complex structure by giving
 the period matrix 
$\Omega$, for which $\Sigma$ has area $\det\Im\Omega$, and which defines the basis $\{\omega_I|I=1,\ldots,g\}$ of Abelian 
differentials and the local complex coordinate $z$ on $\Sigma$. Since we will be considering either Abelian gauge fields at genus
$g$ or non-Abelian gauge fields at genus one, the period matrix will suffice to define the complex structure; we will not need to give
explicit definitions of either Teichm\"uller or moduli space coordinates. We also use the notation $\omega_I=\omega_I(z)dz$, and 
when it can be done unambiguously we keep the index $I$ implicit. The integration measure on $\Sigma$ is normalized to $d^2x=dz\wedge d\bar{z}/(-2i)$, so that 
$\int_\Sigma d^2x \omega_I(z)\overline{\omega}_J(\bar{z})=(\Im \Omega)_{IJ}$. 

One of our goals is to establish the equivalence of three different quantities. The first is a path integral, in which on the final 
surface $\Sigma_R$ we impose a  holomorphic boundary condition 
that fixes the antiholomorphic part $A_{\bar{z}}d\bar{z}$ of the gauge connection $A$, while on the initial surface 
$\Sigma_0$ we fix the component of $A$ along the contractible cycles. The second is a transition amplitude under radial evolution, from an initial state of definite holonomy along the contractible cycles to a coherent final state. The third is a wave function in a 
coherent state basis, obtained by integrating over the gauge orbit a seed wave function which is an eigenstate of the holonomy operator 
along the contractible cycles. 

These quantities will be compared to the Chern-Simons partition functions that are identified with the wave functions obtained by a holomorphic quantization on the Riemann surface $\Sigma$ \cite{Axelrod:1989xt}\cite{Bos:1989kn}\cite{Elitzur:1989nr}. The basis wave functions spanning the gauge-invariant Hilbert space were explicitly given in~\cite{Bos:1989kn} as
\begin{align}
&\quad \mathcal{Z}(A_{\bar{z}},\mu;\Omega)=\frac{e^{+\frac{k\pi}{2}u(\Im \Omega)^{-1} u}}{\widetilde{F}(\Omega)^{\frac{1}{2}}}e^{+\frac{k}{2\pi} \int_\Sigma d^2x \partial_z \chi \partial_{\bar{z}}\chi}\theta\begin{bmatrix}\frac{\mu}{k}\\0\end{bmatrix}(ku,k\Omega)\label{ZZKahler},\\
& A_{\bar{z}}d\bar{z} =\partial_{\bar{z}}\chi d\bar{z}+ i\pi u(\Im\Omega)^{-1}\overline{\omega},\quad \mu\in\mathbb{Z}^g_k,\;k/2\in\mathbb{Z}.\label{Azbar}
\end{align}
The complex number $u$ defines the harmonic part of the differential $A_{\bar{z}}d\bar{z}$ while the integer-valued vector
$\mu$ labels the independent vectors spanning the basis of the Hilbert space. 
Moreover, $k$ is the Chern-Simons level, $\chi$ is a periodic function on $\Sigma$, and $\theta\begin{bmatrix}a\\b\end{bmatrix}(u,\Omega)$ is the Riemann theta function with characteristics \cite{Mumford:1338263}, as defined in \eqref{mumford2}. 
 $\widetilde{F}(\Omega)^{\frac{1}{2}}$ is the ``holomorphic square root" of the scalar Laplace determinant on 
$\Sigma$~\cite{McIntyre:2004xs},
\begin{align}
\frac{\det\nolimits'\Delta}{ \Im \det \Omega}=|\widetilde{F}(\Omega)|^2 \exp(-S_{ZTL}).  \label{ztl}
\end{align}
The obstruction to holomorphic factorization~\cite{quill}, $S_{ZTL}$, is the nonholomorphic part of the Liouville action defined by Zograf and
Takhtajan (see~\cite{McIntyre:2004xs,zo-tak}).
For genus one on the flat metric, $\widetilde{F}(\Omega)^{\frac{1}{2}}$ 
coincides with the Dedekind eta function: $\widetilde{F}(\tau)^{\frac{1}{2}}=\eta(\tau)$.\footnote{At genus one, the Liouville action $S_L$ defined
    in refs.~\cite{McIntyre:2004xs,zo-tak} is related to ours by
    $S_L=S_{ZTL}-\pi i\tau/6 + \pi i \bar{\tau}/6$.}

\subsection{The torus case}
As a warm-up, we first consider the case where $M$ is the solid three-dimensional torus. On each constant-radius surface 
$\Sigma=T^2$ (which is a two-torus) the period matrix is the modular parameter $\tau\equiv\tau_1+i\tau_2$ and defines the global 
holomorphic coordinate $z$ on the torus. From this, we can define local real coordinates $x^{1,2}$ by $z\equiv x^1+ix^2$, where 
$x^1\sim x^1+1$ parametrizes the contractible cycle on $M$. 
The restriction to $T^2$ of a one-form field $A$, $(A_1dx^1+A_2dx^2)$, satisfies
\begin{align}
\begin{cases}
A_1&=\hphantom{i(}A_z+A_{\bar{z}}\\
A_2&=i(A_z-A_{\bar{z}})
\end{cases}\;\Leftrightarrow\;\begin{cases}
A_z&=\frac{1}{2}( A_1-iA_2)\\
A_{\bar{z}}&=\frac{1}{2}( A_1+iA_2)
\end{cases}.\label{A12zzb}
\end{align}
Although these real coordinates are also valid locally on higher-genus Riemann surfaces, for those cases we will use a better description, given in terms of Strebel differentials~\cite{strebel1984}.

In the next subsections, we establish the equivalence between the three quantities mentioned earlier:  the partition function given as
a path integral, the transition amplitude, and the gauge invariant wave function obtained from an appropriate ``seed'' wave function.

\subsubsection*{The path integral}
We impose a holomorphic final condition, fixing $A_{\bar{z}}d\bar{z}|_{\Sigma_R}=\partial_{\bar{z}}\chi d\bar{z}+ i\pi u \tau_2^{-1}\overline{\omega}$, as in \eqref{Azbar}; on the torus, $\overline{\omega}=d\bar{z}$. In addition, as initial condition  we fix
the component of $A$ along the contractible cycle $x^1$ to some $A_1|_{\Sigma_0}=A_1^{(0)}$. The corresponding Chern-Simons partition function $Z(A_{\bar{z}}|_{\Sigma_R},A_1^{(0)};\tau)$ is given by the path integral
\begin{align}
Z(&A_{\bar{z}}|_{\Sigma_R},A_1^{(0)};\tau)\equiv C \int_{\substack{A_{\bar{z}}|_{\Sigma_R}\\A_1|_{\Sigma_0}=A_1^{(0)}}} DA e^{i(I_{\text{CS}}+I_\text{B})},\text{ where }\\
I_{\text{CS}}&=-\frac{k}{4\pi}\int_M AdA-\frac{ik}{2\pi}\int_{\Sigma_R} d^2x A_z A_{\bar{z}},\quad I_\text{B}=+\frac{k}{4\pi} \int_{\Sigma_0}d^2x (A_1A_2+f_{\text{B}}[A_1]).\label{IBAbA1}
\end{align}
In \eqref{IBAbA1}, the boundary term $I_{\text{B}}=(k/4\pi)\int_{\Sigma_0}d^2x (A_1A_2+f_{\text{B}}[A_1])$, with $f_{\text{B}}[A_1]$ an 
arbitrary functional of $A_1$, is an appropriate choice for fixing $A_1$ on the initial surface $\Sigma_0$.
 The normalization constant $C$ may depend on the complex structure and it can be fixed only by imposing additional
 conditions on the partition function. The path integral makes the
the wave function gauge invariant.
An explanation about the gauge invariance of $Z(A_{\bar{z}}|_{\Sigma_R},A_1^{(0)};\tau)$ is in order. We are considering the gauge 
group $U(1)$, not $\mathbb{R}$. The distinction is that $U(1)$ includes large gauge transformations defined on the boundary 
$\Sigma_R$ of the handlebody $M$. A large gauge transformation that has a non-trivial winding along a homotopy cycle of $\Sigma_R$ 
that is contractible in $M$ cannot be extended smoothly to $M$. This implies that the partition function is a sum of terms that are {\em not} related by bulk gauge transformations.

Integrating out $A_r$, the path integral imposes $F_{12}=0$ \cite{Elitzur:1989nr}, so we get
\begin{align}
&\quad Z(A_{\bar{z}}|_{\Sigma_R},A_1^{(0)};\tau)\nonumber\\
&=C\int_{\substack{A_{\bar{z}}|_{\Sigma_R}\\A_1|_{\Sigma_0}=A_1^{(0)}}} DA_1DA_2 \delta(F_{12})\exp{ \left(-\frac{ik}{2\pi}\int_M d^2x dr A_2 \partial_r A_1 \right) }\label{ZCS1}\\
&\qquad \times\exp{\left(-\frac{k}{2\pi} \int_{\Sigma_R}d^2x A_{\bar{z}}A_{\bar{z}} +\frac{k}{\pi}\int_{\Sigma_R}d^2x A_{\bar{z}} A_1 -\frac{k}{4\pi}\int_{\Sigma_R}d^2x A_1 A_1 \right)}\nonumber\\
&\qquad \times \exp \left(+\frac{ik}{4\pi} \int_{\Sigma_0}d^2x f_{\text{B}}[A_1] \right)\nonumber.
\end{align}
The standard procedure is to express $(A_1dx^1+A_2dx^2)$ as a flat connection, resulting in a chiral Wess-Zumino-Witten path integral on the final surface \cite{Elitzur:1989nr}. 

\subsubsection*{The transition amplitude}
We turn now to the coherent state method. The first term in~\eqref{ZCS1} (the bulk term) defines the 
symplectic structure of the theory, implying that $A_1$ and $A_2$ are conjugate variables and satisfy upon quantization 
the equal-radius canonical commutation relation:
\begin{subequations}
\begin{align}
[A_1 (x),A_2(y)]&=-i\left(\frac{2\pi}{k}\right)\delta^{(2)}(x,y)\\
\Leftrightarrow \; [A_{\bar{z}}(x),A_z(y)]&=-\frac{\pi}{k}\delta^{(2)}(x,y).\label{torusAzAzbarcomm}
\end{align}
\end{subequations}
Here $\delta^{(2)}(x,y)$ denotes the delta function with respect to the $(x^1,x^2)$-coordinates. Moreover, define the $A_1$-eigenstate $|A_1\rangle$ as a translation from the $A_1=0$ eigenstate $|0\rangle$ 
\textit{effected by applying the conjugate momentum $\hat{A}_2$}
\begin{align}
|A_1\rangle&\equiv \bar{C} \exp \left(+\frac{ik}{2\pi}\int_\Sigma d^2x \hat{A}_2 A_1 \right)|0\rangle.\label{A1translate0}
\end{align}
Here too $C$ is a normalization constant, which we leave arbitrary for the time being. Using~\eqref{A1translate0} together with \eqref{A12zzb}, we can construct the wave function of the coherent state $|A_z)$ in the 
$|A_1\rangle$-basis,
\begin{align}
\langle A_1 |A_z )&= C \exp{\left( -\frac{k}{2\pi}\int_\Sigma d^2x A_z^2+\frac{k}{\pi}\int_\Sigma d^2x A_z A_1 -\frac{k}{4\pi} \int_\Sigma d^2x A_1^2 \right)},\label{cohalA1}
\end{align}
which satisfies the defining properties (with $A_{\bar{z}}=A_{z}^*$),
\begin{subequations}
\begin{empheq}[left=\empheqlbrace]{align}
\hat{A}_z(x)\langle A_1 |A_z )&=\frac{1}{2}\left(A_1(x)+\frac{2\pi}{k}\frac{\delta}{\delta A_1(x)} \right)\langle A_1 |A_z )=A_z(x)\langle A_1 |A_z ),\\
\hat{A}_1(x)(A_{\bar{z}}|A_1\rangle&=\left(+\frac{\pi}{k}\frac{\delta}{\delta A_{\bar{z}}(x)} +A_{\bar{z}}(x)\right)  (A_{\bar{z}}|A_1\rangle=A_1(x)(A_{\bar{z}}|A_1\rangle.
\end{empheq}
\end{subequations}
Let us consider the transition amplitude, from an $A_1$-eigenstate $|A_1^{(0)}\rangle$ on the initial surface $\Sigma_0$, to a coherent state $|A_z^R)$ on the final surface $\Sigma_R$, as we radially evolve the system with the Hamiltonian read off from \eqref{ZCS1}:
\begin{align}
&\quad( A_{\bar{z}}^R | e^{-i HR}| A_1^{(0)}\rangle= \int DA_1^R (A_{\bar{z}}^R |A_1^R\rangle \langle A_1^R|e^{-i HR}| A_1^{(0)}\rangle\label{amp1}\\
&=C \int DA_1^R\exp{\left( -\frac{k}{2\pi}\int_{\Sigma_R} d^2x{A_{\bar{z}}^R}^2+\frac{k}{\pi}\int_{\Sigma_R} d^2x A_{\bar{z}}^R A_1^R -\frac{k}{4\pi} \int_{\Sigma_R} d^2x (A_1^R)^2 \right)}\\
&\qquad \times \int_{\substack{A_1|_{\Sigma_R}=A_1^R\\A_1|_{\Sigma_R}=A_1^{(0)}}} DA_1DA_2 \delta(F_{12}) \exp{\left( -\frac{ik}{2\pi}\int_M d^2x dr A_2\partial_r A_1\right)}\nonumber\\
&=C \int_{A_1|_{\Sigma_R}=A_1^{(0)}}DA_1DA_2 \delta(F_{12}) \exp{\left( -\frac{ik}{2\pi}\int_M d^2x dr A_2\partial_r A_1\right)}\label{amp2}\\
&\qquad\qquad\qquad \times  \exp{\left( -\frac{k}{2\pi}\int_{\Sigma_R} d^2x{A_{\bar{z}}^R}^2+\frac{k}{\pi}\int_{\Sigma_R} d^2x A_{\bar{z}}^R A_1^R -\frac{k}{4\pi} \int_{\Sigma_R} d^2x (A_1^R)^2 \right)} .\nonumber
\end{align}
This is identical to the partition function \eqref{ZCS1}, $Z(A_{\bar{z}}|_{\Sigma_R},A_1^{(0)};\tau)$, with the boundary term $f_{\text{B}}[A_1]=0$, and $A_{\bar{z}}^R=A_{\bar{z}}|_{\Sigma_R}$. In both cases, we have imposed the initial condition 
$A_1|_{\Sigma_0}=A_1^{(0)}$. The equivalence between \eqref{ZCS1} and~\eqref{amp2} holds for arbitrary genus 
because it only relies on a local decomposition of the complex coordinate $z$ into real coordinates that is independent of the topology of
the surface $\Sigma$. From now on, without ambiguity, we drop the superscript $R$ from $A_{\bar{z}}^R$. 

Next, we evaluate Eq.~\eqref{amp2} and find out what it computes for the torus case. We parametrize the $A_{1,2}$ that solve the constraint $F_{12}=0$ by
\beq
\begin{cases}
A_1(r,x^1,x^2)&=a_1(r)+2\pi n+\partial_1 \lambda_0(r,x^1,x^2)\\
A_2(r,x^1,x^2)&=a_2(r)+\partial_2 \lambda_0(r,x^1,x^2),
\end{cases}
\eeq{cases}
where $\lambda_0(r,x^1,x^2)$ is a periodic function on $\Sigma$, and the shift in $A_1$ by $2\pi n$ with $n\in\mathbb{Z}$ comes from 
the large gauge transformations that are singular inside the bulk. Note that a shift in $\lambda_0(r,x^1,x^2)$ by any $x^1$-independent function $f_2(r,x^2)$ 
also solves $F_{12}=0$ and leaves the integrand of the path integral invariant, thus the $x^1$-independent modes can be factored out
 of the path integral and consistently discarded\footnote{This can be shown more rigorously using the BRS formalism~\cite{oblak}.}. 
On the other hand, shifting $\lambda_0(r,x^1,x^2)$ by some $f_1(r,x^1)$ changes the boundary action, so these modes cannot be 
factored out from the path integral. We restrict our initial condition to $A_1|_{\Sigma_0}=a_1(0)$ with $a_1(0)=$ constant--- this is a 
natural choice since the initial surface is in fact degenerate, so $\lambda_0(r=0,x^1,x^2)=\lambda_0(r=0,x^2)$ is independent of $x^1$.
 The integration measure in \eqref{amp2} satisfies \cite{Elitzur:1989nr}
\begin{align}
DA_1DA_2 \delta(F_{12})=Da_1Da_2D'\lambda_0,
\end{align}
i.e. the change of variables~\eqref{cases} has unit Jacobian. Here a prime denotes discarding $x^1$-independent functions. Moreover, as in \eqref{Azbar}, $A_{\bar{z}}=\partial_{\bar{z}}\chi + i\pi u\tau_2^{-1}$. The amplitude \eqref{amp2} becomes
\begin{align}
&\quad( A_{\bar{z}} | e^{-i HR}| a_1(0)\rangle\nonumber\\
&=C e^{+\frac{k\pi}{2}u\tau_2^{-1}u+\frac{k}{2\pi}\int_{\Sigma_R} d^2x \partial_z\chi\partial_{\bar{z}}\chi }\times\int D'\widetilde{\lambda}_0 \exp\left(-\frac{k}{2\pi}\int_{\Sigma_R}d^2x \partial_1\widetilde{\lambda}_0 \partial_{\bar{z}} \widetilde{\lambda}_0 \right)\label{A1torustrial2}\\
&\quad \times  \sum_{n\in\mathbb{Z}}\exp{\left(+ik\pi u (a_1(0)+2\pi n)-\frac{k}{4\pi}(a_1(0)+2\pi n)\tau_2 (a_1(0)+2\pi n)\right)} \nonumber\\
&=\frac{e^{+\frac{k\pi}{2}u\tau_2^{-1} u}}{\eta(\tau)}e^{+\frac{k}{2\pi}\int_{\Sigma_R} d^2x \partial_z\chi\partial_{\bar{z}}\chi }
\theta\begin{bmatrix}\frac{a_1(0)}{2\pi}\\0\end{bmatrix}(ku,ik\tau_2). \label{ampl-m1}
\end{align}
In arriving at \eqref{A1torustrial2}, we integrated out $a_2(r)$ to obtain an $r$-independent $a_1(r)=a_1$. Together with the initial 
condition $A_1(r=0)=a_1(0)$, this means $a_1(R)=a_1(0)$. We also defined $\widetilde{\lambda}_0\equiv \lambda_0+\chi$. 

The path integral on $\widetilde{\lambda}_0$ equals $\det^{-1/2}({-\frac{k}{2\pi}} \partial_{\bar{z}} \partial_1)$. We are still free to 
choose the constant $C$. Besides removing ultraviolet divergences in the functional determinant, it can be further fixed by requiring that eq.~\eqref{ampl-m1} be a section of a projectively flat connection on the moduli space of complex 
structures~\cite{Axelrod:1989xt}. This is simply the requirement that the scalar product of the base wave functions~\eqref{ampl-m1} 
must be independent of the complex structure. By making this choice we get
$1/\widetilde{F}(\Omega)^{\frac{1}{2}}=1/\eta(\tau)$.~\footnote{In the genus-1 case $C$ is in fact independent of the complex structure,
as can be seen by an explicit computation~\cite{oblak}.}
Let us compare this to the wave
 functions $\mathcal{Z}(A_{\bar{z}},\mu;\Omega)$ \eqref{ZZKahler} obtained from holomorphic quantization, that span 
 the gauge-invariant Hilbert space. On the genus $g=1$ torus $\Sigma=T^2$, they are given by
\begin{align}
\mathcal{Z}(A_{\bar{z}},\mu;\tau)&=\frac{e^{+\frac{k\pi}{2}u \tau_2^{-1} u}}{\eta(\tau)}e^{+\frac{k}{2\pi} \int_\Sigma d^2x \partial_z \chi \partial_{\bar{z}}\chi}\theta\begin{bmatrix}\frac{\mu}{k}\\0\end{bmatrix}(ku,k\tau)\label{ZZKahler2},\\
&\text{where }A_{\bar{z}}=\partial_{\bar{z}}\chi + i\pi u\tau_2^{-1},\text{ and }\mu=0,1,\ldots,k-1.\nonumber
\end{align}
As we set $a_1(0)/2\pi=\mu/k$, we see that the wave function $(A_{\bar{z}} | e^{-i HR}| a_1(0)\rangle$, or equivalently the path integral 
$Z(A_{\bar{z}}|_{\Sigma_R},A_1^{(0)};\tau)$ in \eqref{ZCS1} with $f_{\text{B}}[A_1]=0$, differs by a $\tau_1$-dependent 
phase in the theta function from the function
$\mathcal{Z}(A_{\bar{z}},\mu;\tau)$ in~\eqref{ZZKahler2}. So, the wave functions $(A_{\bar{z}} | e^{-i HR}| a_1(0)\rangle$ with $a_1(0)/2\pi=\mu/k$, $\mu=0,1,\ldots,k-1$ exhibit a dependence on the complex structure different from that of the basis wave functions \eqref{ZZKahler2}.

We cannot reabsorb this difference into a redefinition of the constant $C$ without giving up one of the objectives of our paper, which is to establish a state-operator correspondence associating each state obtained by applying Wilson loops
to the vacuum to the partition function of Chern-Simons on a solid torus containing the same Wilson loop. So, once we normalize the vacuum and the vacuum partition function, we cannot further normalize separately the 
other partition functions. What we can do is to understand where the discrepancy comes from and try to fix it by appropriately changing
the definition {\em of the Wilson loop operator}.

To find the meaning of this discrepancy, we consider a different basis on the torus. We define global coordinates $(\phi,t)$ which both
have unit period, so that $z=\phi+\tau t$, $\phi \sim \phi +1$, $\tau \sim \tau +1$ and
\begin{align}
\begin{cases}
A_\phi&=\hphantom{\tau} A_z+\hphantom{\tau}A_{\bar{z}}\\
A_t&=\tau A_z+ \bar{\tau} A_{\bar{z}}
\end{cases}\;\Leftrightarrow\;\begin{cases}
A_z&=(\tau-\bar{\tau})^{-1}(-\bar{\tau} A_{\phi}+A_t)\\
A_{\bar{z}}&=(\tau-\bar{\tau})^{-1}(\hphantom{-}\tau A_{\phi}-A_t)
\end{cases}.\label{Aphitzzb}
\end{align}
In particular, $A_\phi=A_1$, but $A_t=\tau_2 A_2+\tau_1 A_1\neq A_2$ in general. The conjugate variables $(A_\phi=A_1,\tau_2^{-1}A_t)$ are related to the previous conjugate variables $(A_1,A_2)$ by a canonical transformation which simply shifts $A_2$ by a term linear in $A_1$. The canonical commutation relation is
\begin{subequations}
\begin{align}
[A_\phi (x),A_t(y)]&=-i \tau_2 \left(\frac{2\pi }{k}\right)\delta^{(2)}(x,y)\\
\Leftrightarrow\; [A_{\bar{z}}(x),A_z(y)]&=-\frac{\pi}{k}\delta^{(2)}(x,y).
\end{align}
\end{subequations}
Here $\delta^{(2)}(x,y)$ is again the delta function in the $(x^{1},x^2)$ coordinates. 
Similarly to \eqref{A1translate0}, we define the $A_\phi$-eigenstate $|A_\phi\rangle\rangle$ by translating 
the $A_\phi=0$ eigenstate $|0\rangle\rangle\equiv |0\rangle$, but this time with the operator
 $\hat{A}_t$,
 \begin{align}
|A_\phi\rangle\rangle&\equiv \bar{C}\exp{\left(+\frac{ik}{2\pi}\tau_2^{-1}\int_\Sigma d^2x \hat{A}_t A_\phi \right) }|0\rangle\rangle\text{ and }|0\rangle\rangle\equiv |0\rangle.\label{Aphitranslate0}
\end{align} 
The eigenstates $|A_\phi\rangle\rangle$ and $|A_\phi\rangle$ are related by a pure phase, 
\begin{align}
|A_\phi\rangle\rangle&=\exp{\left( +\frac{ik}{4\pi}\tau_1 \tau_2^{-1} \int_\Sigma d^2x A_\phi^2\right)}|A_\phi\rangle.\label{Aphi1phase0}
\end{align}
By using \eqref{Aphitzzb}, we see that the wave function of the coherent state $|A_z)$ in the $|A_\phi\rangle\rangle$-basis 
differs from~\eqref{cohalA1} 
\begin{align}
\langle\langle A_\phi |A_z )&=C \exp{\left( -\frac{k}{2\pi}\int_\Sigma d^2xA_z^2+\frac{k}{\pi}\int_\Sigma d^2x A_z A_\phi -\frac{ik }{4\pi }\bar{\tau}\tau_2^{-1} \int_\Sigma d^2x A_\phi^2 \right)}.\label{cohalAphi}
\end{align}
Repeating the same calculations as above, one finds that
\begin{align}
&\quad( A_{\bar{z}} | e^{-i HR}| a_\phi(0)\rangle\rangle\nonumber\\
&=C e^{+\frac{k\pi}{2}u\tau_2^{-1} u+\frac{k}{2\pi}\int_{\Sigma_R} d^2x \partial_z\chi\partial_{\bar{z}}\chi }\times\int D'\widetilde{\lambda}_0 \exp\left(-\frac{k}{2\pi}\int_{\Sigma_R}d^2x \partial_1\widetilde{\lambda}_0 \partial_{\bar{z}} \widetilde{\lambda}_0 \right)\\
&\qquad\times\sum_{n\in\mathbb{Z}}\exp{\left(iku (a_\phi(0)+2\pi n)+\frac{ik}{4\pi}(a_\phi(0)+2\pi n) \tau (a_\phi(0)+2\pi n)\right)}\nonumber\\
&=\frac{e^{+\frac{k\pi}{2}u\tau_2^{-1} u}}{\eta(\tau)}e^{+\frac{k}{2\pi}\int_{\Sigma_R} d^2x \partial_z\chi\partial_{\bar{z}}\chi }\theta\begin{bmatrix}\frac{a_\phi(0)}{2\pi}\\0\end{bmatrix}(ku,k\tau), \label{A1torus2}
\end{align}
where we normalized $C$ as in Eq.~\eqref{ampl-m1}. 
This \textit{is} exactly one of the \eqref{ZZKahler2} when we set $a_1(0)/2\pi=\mu/k$. Thus, we learn that to get an answer holomorphic in the complex structure $\tau$ we need a particular choice of canonical variables $(A_\phi,A_t)$, or equivalently a particular choice of eigenstate
$|A_\phi\rangle\rangle$. In terms of the path integral $Z(A_{\bar{z}}|_{\Sigma_R},a_\phi(0);\tau)$, this corresponds to a particular choice of the boundary term, namely: $f_{\text{B}}[A_1]=\tau_1 \tau_2^{-1} A_1^2$.

\subsubsection*{The gauge-invariant wave function}
Under a gauge transformation
\begin{align}
A_{\bar{z}}\longrightarrow {}^\lambda  A_{\bar{z}}&\equiv A_{\bar{z}}-\partial_{\bar{z}}\lambda,
\end{align}
a holomorphic wave function $\Psi[A_{\bar{z}}]$ obtained  from  K\"ahler quantization transforms as \cite{Elitzur:1989nr,Bos:1989kn}
\begin{align}
\Psi[A_{\bar{z}}]\longrightarrow (U(\lambda)\cdot \Psi)[A_{\bar{z}}]&\equiv \exp{\left(- \frac{k}{2\pi}\int_\Sigma d^2x \partial_z \lambda \partial_{\bar{z}}\lambda+\frac{k}{\pi}\int_\Sigma d^2x \partial_z \lambda A_{\bar{z}} \right)}\Psi[{}^\lambda A_{\bar{z}}].
\end{align}
Here $\lambda$ can include large gauge transformations. Thus, starting from any ``seed" wave function $\Psi_0[A_{\bar{z}}]$ we can integrate over the gauge group to construct a gauge-invariant wave function:
\begin{align}
\Psi[A_{\bar{z}}]&\equiv \int D'\lambda (U(\lambda)\cdot \Psi_0)[A_{\bar{z}}]. \label{UPsi0torus}
\end{align}
This formula includes a sum over large gauge transformations, so the most general $\lambda$ is 
\begin{align}
\lambda&\equiv \lambda_0+\lambda' ,
\end{align}
where $\lambda_0$ is periodic on the torus, while the multivalued large gauge parameter $\lambda'$ enters in the integral only through its derivatives, which are single-valued on the torus; they are given by
\begin{equation}
\begin{cases}
\partial_{\bar{z}}\lambda'&=+i\pi (m+n\tau)\tau_2^{-1}\\
\partial_z\lambda'&= -i\pi (m+n\overline{\tau})\tau_2^{-1}
\end{cases},\;m,n\in\mathbb{Z}.
\end{equation}   
If we take $(A_{\bar{z}}|a_1(0)\rangle\rangle$ to be a seed wave function 
$\Psi_0[A_{\bar{z}}]$ which is not necessarily gauge-invariant and integrate over all gauge transformations including the large transformations $m,n\in\mathbb{Z}$, we reproduce the theta function in \eqref{ZZKahler2}.
To see this, we impose again the conditions \eqref{Azbar} $A_{\bar{z}} =\partial_{\bar{z}}\chi+ i\pi u\tau_2^{-1}\equiv \partial_{\bar{z}}\chi+\partial_{\bar{z}}\chi'$ and $a_\phi(0)=2\pi\mu/k$, $\mu\in\mathbb{Z}$ and find
\begin{align}
&\quad \int D'\lambda (U(\lambda)\cdot \Psi_0)[A_{\bar{z}}]\nonumber\\
&=\sum_{m,n\in\mathbb{Z}} \int D'\lambda_0 \exp{\left(- \frac{k}{2\pi}\int_\Sigma d^2x \partial_z \lambda \partial_{\bar{z}}\lambda+\frac{k}{\pi}\int_\Sigma d^2x\partial_z \lambda A_{\bar{z}} \right)}({}^{\lambda} A_{\bar{z}}|a_1(0)\rangle\rangle\\
&=\sum_{m,n\in\mathbb{Z}} C\int D'\lambda_0 \exp{\left(-\frac{k}{2\pi}\int_\Sigma d^2x \partial_1 \lambda_0 \partial_{\bar{z}}\lambda_0+\frac{k}{\pi}\int_\Sigma d^2x A_{\bar{z}} \partial_1 \lambda_0 \right)}\label{torusgaugeint}\\
&\quad \times\exp{\left(-\frac{k}{2\pi}\int_\Sigma d^2x A_{\bar{z}}^2-\frac{k}{2\pi}\int_\Sigma d^2x \partial_1 \lambda' \partial_{\bar{z}}\lambda'+\frac{k}{\pi}\int_\Sigma d^2x A_{\bar{z}} \partial_1 \lambda'   \right)}\nonumber\\
&\quad \times \exp \left(+\frac{k}{\pi}\int_\Sigma d^2x A_{\bar{z}} a_1(0)-\frac{k}{\pi}\int_\Sigma d^2x\partial_{\bar{z}}\lambda' a_1(0)+\frac{ik }{4\pi }\tau\tau_2^{-1} \int_\Sigma d^2x (a_1(0))^2\right)\nonumber\\
&=\sum_{m,n\in\mathbb{Z}} C\int D'\lambda_0 \exp{\left(-\frac{k}{2\pi}\int_\Sigma d^2x \partial_1 \lambda_0 \partial_{\bar{z}}\lambda_0+\frac{k}{\pi}\int_\Sigma d^2x A_{\bar{z}} \partial_1 \lambda_0 \right)}\\
&\quad \times\exp{\left(-\frac{k}{2\pi}\int_\Sigma d^2x A_{\bar{z}}^2 +ik\pi nm-2\pi ikm \left( \frac{a_1(0)}{2\pi} \right)\right)}\nonumber\\
&\quad \times \exp \left(+2k\int_\Sigma d^2x A_{\bar{z}}\left(\frac{a_1(0)}{2\pi}-n\right) +i\pi k\left(\frac{a_1(0)}{2\pi}-n\right)\tau \left(\frac{a_1(0)}{2\pi}-n\right) \right)\nonumber\\
&=C\exp{\left(+\frac{k}{2\pi}\int_\Sigma d^2x \partial_z \chi\partial_{\bar{z}} \chi+\frac{k\pi}{2}u \tau_2^{-1} u \right)}\int D'\tilde{\lambda}_0 \exp{\left(-\frac{k}{2\pi}\int_\Sigma d^2x \partial_1 \tilde{\lambda}_0 \partial_{\bar{z}}\tilde{\lambda}_0\right)}\label{nom1}\\
&\times \sum_{m,n\in\mathbb{Z}}\exp \left(ik\pi n\tau n+2\pi i k u\left(\frac{a_1(0)}{2\pi}-n\right) +i\pi k\left(\frac{a_1(0)}{2\pi}-n\right)\tau \left(\frac{a_1(0)}{2\pi}-n\right) \right)\nonumber\\
&= e^{+\frac{k}{2\pi}\int_\Sigma d^2x \partial_z \chi\partial_{\bar{z}} \chi}
\frac{e^{+\frac{k\pi}{2}u \tau_2^{-1} u}}{\eta(\tau)}\theta\begin{bmatrix}
\frac{\mu}{k}\\0
\end{bmatrix}(ku,k\tau),
\end{align}
which is exactly $\mathcal{Z}(A_{\bar{z}},\mu;\tau)$  in \eqref{ZZKahler2}. In arriving at \eqref{nom1}, 
the constant $C$ was fixed as in~\eqref{ampl-m1} using the normalization condition $C\det^{-1/2}(-{\frac{k}{2\pi}} \partial_{\bar{z}}\partial_1) =\eta(\tau)^{-1}$ 
and we used 
that $k/2\in\mathbb{Z}_{>0}$, $m\in\mathbb{Z}$ and $ka_1(0)/2\pi=\mu\in\mathbb{Z}$, so the summand does not depend on $m$, 

Because of this, in the last line we discarded the infinite sum over $m\in\mathbb{Z}$. Similarly, the same calculation done
with  $(A_{\bar{z}}|a_1(0)\rangle$ as seed wave function reproduces \eqref{A1torustrial2}. Notice that discarding the sum over $m$ means simply to remove identical gauge copies from the definition of the 
gauge-invariant wave function. This is a standard part of the construction of a gauge invariant, normalizable state or operator 
using an integral (and/or sum) over
gauge transformations. Its analog in the
context of three-dimensional gravity is explained for instance in~\cite{Maloney:2007ud}.

\subsubsection{Blowing up Wilson loops}\label{WLtorusPsi}
One can insert into the path integral a gauge-invariant Wilson loop operator defined along a loop $C$ on $M$, as
\begin{align}
\hat{W}_\mu[C]&\equiv \mathcal{P}\exp{\left(i\mu \oint_C \hat{A} \right)},\;\mu\in\mathbb{Z}.
\end{align}
$\mathcal{P}$ means path-ordering, and the $U(1)$ charge $\mu$ is integer-valued such that $\hat{W}_\mu[C]$ is invariant under large
 gauge transformations defined on $C$. We restrict $C$ to be a path that runs along the non-contractible cycle of $M$, and without  
 loss of generality put it at the origin $r=0$ of the solid torus. 

We would like to map $\hat{W}_\mu[C]$ to a ``blown-up" operator in radial quantization, which acts on a 
state defined on $\Sigma$. To this 
end, recall that the $A_1$-eigenstate $|A_1\rangle$ is the translation of the $A_1=0$ eigenstate $|0\rangle$ by the operator 
$\hat{A}_2$ given in \eqref{A1translate0},
\begin{align}
|A_1\rangle&\equiv\exp{\left(+\frac{ik}{2\pi}\int_\Sigma d^2x \hat{A}_2 A_1 \right) }|0\rangle.\label{A1translate}
\end{align}
The initial surface $\Sigma_0$ is degenerate but the Wilson loop operator $\hat{W}_\mu[C_2]$, with $C_2$ at $r=0$ running along the $x^2$-direction, can be ``blown-up" and identified with the translation operator defined in~\eqref{A1translate} acting on the Hilbert space on $\Sigma$,
\begin{align}
\hat{W}_\mu[C_2]&=\mathcal{P}\exp{\left(i\mu \oint dx^2 \hat{A}_2 \right)}\label{WC20}\\
\longrightarrow \hat{W}_{\mu}[\Sigma,2] &\equiv \exp{\left(i\int_\Sigma d^2x \mu\hat{A}_2  \right) }=\exp{\left(i\int_\Sigma d^2x \left(i\mu\hat{A}_z - i\mu \hat{A}_{\bar{z}}\right)\right) }.\label{WC2}
\end{align}
Alternatively, choosing $A_t$ as the conjugate momentum from~\eqref{Aphitranslate0} we have
\begin{align}
|A_\phi\rangle\rangle&\equiv\exp{\left(+\frac{ik}{2\pi }\tau_2^{-1}\int_\Sigma d^2x \hat{A}_t A_\phi \right) }|0\rangle.\label{Aphitranslate}
\end{align}
We can also define a ``blown-up" version of the Wilson loop operator $\hat{W}_\mu[C_t]$, with $C_t$ at $r=0$ running along the $t$-direction, and identify it with the translation operator in \eqref{Aphitranslate},
\begin{align}
\hat{W}_\mu[C_t]&=\mathcal{P}\exp{\left(i\mu \int_0^1 dt \hat{A}_t \right)}\\
\longrightarrow \hat{W}_{\mu}[\Sigma,t] &\equiv\exp{\left(i\tau_2^{-1}\int_\Sigma d^2x \mu\hat{A}_t  \right) }=\exp{\left(i\int_\Sigma d^2x \left(\tau \tau_2^{-1}\mu \hat{A}_z +\bar{\tau} \tau_2^{-1}\mu \hat{A}_{\bar{z}}\right) \right) }.\label{WCt}
\end{align}

\subsubsection*{Gauge invariance and framing}
Both $C_2$ and $C_t$ trace the same closed loop at the origin, though with twists differing by $\tau_1$. One may wish to assign a framing to this loop by defining a vector field on it \cite{Witten:1988hf,Witten:1989wf}, thereby extending this loop into a ribbon. Such a vector field must be periodic under the global identification $(x^1,x^2)\sim (x^1+\tau_1,x^2+\tau_2)$--- now that we are away from the degenerate $r=0$ surface. The simplest choice is that corresponding to $C_t$, while that corresponding to $C_2$ does not respect the periodicity.

In the language of the ``blowing-up" procedure, this fact translates to demanding that the ``blown-up" Wilson loop operator on $\Sigma$ must be gauge-invariant. Both of the original Wilson loops $\hat{W}_\mu[C_2]$ and $\hat{W}_\mu[C_t]$ are invariant under gauge transformations defined on the loops. But among their ``blown-up" versions, only $\hat{W}_{\mu}[\Sigma,t]$ is gauge-invariant on $\Sigma$. $\hat{W}_{\mu}[\Sigma,2]$ is not invariant under large transformations with a non-trivial winding along the $\phi$-direction, which were not well-defined transformations on the original loop $C_2$. Gauge invariance thus selects $\hat{W}_{\mu}[\Sigma,t]$ as the preferred operator on $\Sigma$. 

Note that the $\hat{W}_{\mu}[\Sigma,t]$, $\mu\in\mathbb{Z}$, are not the only gauge-invariant operators. The most general gauge invariant ``blown-up" operator on $\Sigma$ with constant coefficients\footnote{We will drop this restriction in the higher-genus cases.} takes the form
\begin{align}
\hat{W}[\Sigma;\mu,N]&\equiv \exp\left(i\mu\int_\Sigma d^2x \left(\tau \tau_2^{-1} \hat{A}_z +\bar{\tau} \tau_2^{-1} \hat{A}_{\bar{z}}\right)+iN \tau_2^{-1}\int_\Sigma d^2x \left(\hat{A}_z + \hat{A}_{\bar{z}}\right) \right)\label{torusWmuN}\\
&=\hat{W}[\Sigma,t]\exp\left(iN \int_\Sigma d^2x \hat{A}_\phi\right)  \exp\left(+\frac{i\pi}{k}\mu N \right),\;\mu,N\in\mathbb{Z},
\end{align}
where we used a special case of the Baker-Campbell-Hausdorff formula:
\begin{align}
\text{if }[X,[X,Y]]&=0=[Y,[X,Y]],\quad\text{ then }\quad e^{X+Y}=e^{X}e^Y e^{-\frac{1}{2}[X,Y]},\label{BCH}
\end{align}
and the canonical commutation relation \eqref{torusAzAzbarcomm}
to arrive at this result. That is, $\hat{W}[\Sigma;\mu,N]$ is equivalent to blowing up the loop $C_t$ together with the loop $C_\phi$ 
along the $\phi$-direction. So we can identify the pure phase as due to the linking of these two loops \cite{Witten:1988hf}. 
Since $\hat{W}[\Sigma;\mu,N]$ is gauge-invariant, we can commute it with the integral over gauge 
transformations~\eqref{UPsi0torus}, and let it act on the seed wave function $\Psi_0[A_{\bar{z}}]=(A_{\bar{z}}|0\rangle\rangle$, which is an eigenvector of $\hat{A}_\phi$  with 
eigenvalue $A_\phi=0$. The result reads
\begin{align}
\hat{W}[\Sigma;\mu,N](A_{\bar{z}}|0\rangle\rangle&=e^{+\frac{i\pi}{k}\mu N}\hat{W}[\Sigma,t] (A_{\bar{z}}|0\rangle\rangle=e^{+\frac{i\pi}{k}\mu N}(A_{\bar{z}}|2\pi\mu/k\rangle\rangle.
\end{align}
The integral \eqref{UPsi0torus} is then
\begin{align}
&\quad \int D'\lambda \left(U(\lambda)\cdot \hat{W}[\Sigma;\mu,N](A_{\bar{z}}|0\rangle\rangle \right)=e^{+\frac{i\pi}{k}\mu N}\int D'\lambda \left(U(\lambda)\cdot (A_{\bar{z}}|2\pi\mu/k\rangle\rangle \right)\\
& = e^{+\frac{i\pi}{k}\mu N}\frac{e^{+\frac{k\pi}{2}u \tau_2^{-1} u}}{\eta(\tau)}e^{+\frac{k}{2\pi}\int_\Sigma d^2x \partial_z \chi\partial_{\bar{z}} \chi}\theta\begin{bmatrix}
\frac{\mu}{k}\\0
\end{bmatrix}(ku,k\tau).
\end{align}
Thus, the only contribution of the  $N$-dependent term in $\hat{W}[\Sigma;\mu,N]$ to the gauge-invariant wave function is a pure phase
 proportional to the linking number. This is expected from general principles of canonical quantization, because the Wilson loops 
 $\hat{W}_{\mu}[C_t]$, $\mu\in\mathbb{Z}_k$, already span the full Hilbert space, and so the Wilson loop operators along $C_\phi$ 
 can at most contribute a pure phase \cite{Polychronakos:1989cd}. Note that in the special case $N=\mu N'$, $N'\in\mathbb{Z}$, $\hat{W}[\Sigma;\mu,N=\mu N']$ can be regarded as the blowing-up of the loop $C_t$ with additional $N'$ twists. The phase 
\begin{align}
\exp\left(+\frac{i\pi}{k}\mu N \right) =\exp\left( +i\pi N'\frac{\mu^2}{k}\right)
\end{align}
is then identified with the framing anomaly \cite{Witten:1988hf}. 


\subsection{Higher genus}
For partition functions on higher-genus handlebodies, it is convenient to make use of certain special 
quadratic differentials on Riemann surfaces, 
reviewed in \hyperref[secquaddiff]{Appendix \ref*{secquaddiff}}. Specifically, we pick a Strebel differential $\varphi$, which is 
a quadratic differential on the Riemann surface $\Sigma$, holomorphic in the complex structure; locally, $\varphi=h(z)dz^2$ where 
$h(z)$ is holomorphic. The existence of such differentials is proven in~\cite{strebel1984}. We do not need to know their precise form. 
All we need from a Strebel differential is the fact that it foliates the Riemann surface $\Sigma$ into horizontal trajectories, which are closed curves given by
\begin{align}
f(p)&\equiv \int^p \sqrt{h}dz-\int^p \sqrt{h^*}d\bar{z}=\text{constant}.
\end{align}
The Strebel differential $\varphi$ also defines a metric on $\Sigma$, which takes the form
\begin{align}
g_{z\bar{z}}=\sqrt{hh^*}=\sqrt{g}.\label{metric2}
\end{align}
This metric may have zeros or singularities, which define the singular points of the foliation.
We define next a vector field $\mathrm{v}$ of unit norm with respect to \eqref{metric2}, whose integral curves are the horizontal trajectories so that $\mathrm{v}(f)\equiv 0$:
\begin{align}
\mathrm{v}&\equiv v(z)\partial_z+\bar{v}(\bar{z})\partial_{\bar{z}}= \frac{1}{\sqrt{h}}\partial_z+\frac{1}{\sqrt{h^*}}\partial_{\bar{z}}\equiv \partial_h.\label{vmetric}
\end{align}
We use the vector field $\mathrm{v}$ to define cycles on a higher-genus Riemann surface $\Sigma$, that are contractible on 
the corresponding handlebody $M$. On the torus, $v=\bar{v}=1$ and $\partial_h=\partial_1$. The square
root in \eqref{vmetric} can cause generically an obstruction to defining a global {\em holomorphic} vector field on $\Sigma$. On the other 
hand, we do not need $\mathrm{v}$ to be holomorphic, so we can always rescale $\mathrm{v}$ by a common factor: 
$\mathrm{v}\rightarrow \epsilon \mathrm{v}$ with
$\epsilon$ a smooth {\em real} function. The equations that we will find in the next subsections depend only on the ratio $\bar{v}/v$,
which is not affected by the rescaling. The function $\epsilon$ can even vanish on subsets of measure zero that are transverse to the 
horizontal trajectories without altering the ratio $\bar{v}/v$. By making $\epsilon$ vanish somewhere on $\Sigma$, a 
{\em nonholomorphic} vector field can be defined everywhere on $\Sigma$.

\begin{figure}
\begin{center}
\includegraphics[width=.8\linewidth,keepaspectratio]{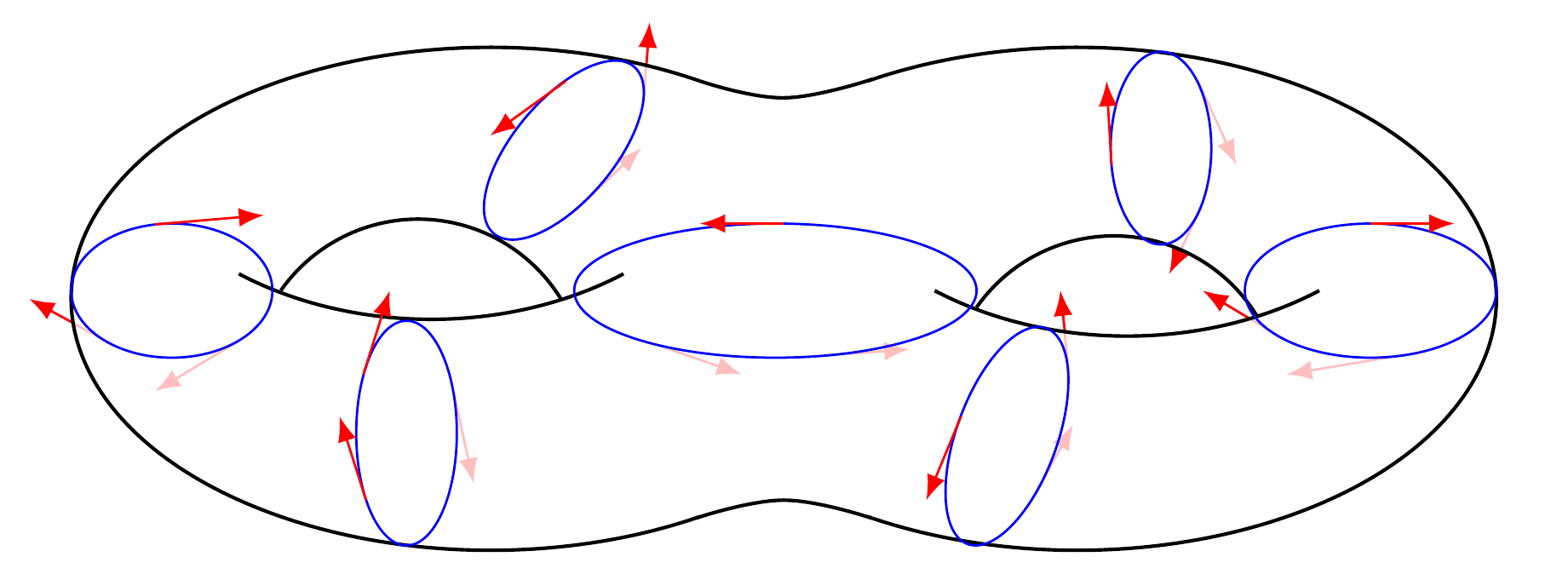}
\caption{A schematic illustration of a Strebel differential on a genus-two Riemann surface, which defines horizontal trajectories denoted by blue loops. The vector field $\mathrm{v}$ that generates the horizontal trajectories is denoted by red arrows.}\label{strebelfig}
\end{center}
\end{figure}

The horizontal trajectories and the vector field $\mathrm{v}$ that generates them are illustrated schematically 
in \hyperref[strebelfig]{Figure \ref*{strebelfig}}.

\subsubsection{The vacuum partition function}

We would like to generalize the third approach used in the torus case: start from a non-gauge invariant seed wave function 
$\Psi_0[A_{\bar{z}}]$ and integrate over the gauge orbit to arrive at a gauge-invariant wave function. The most natural and simplest seed
 wave function is the eigenstate of trivial holonomy along the contractible cycles, $\Psi_0[A_{\bar{z}}]$, computed in the coherent 
 state basis $(A_{\bar{z}}|$. In terms of the vector field $\mathrm{v}$ that generates the closed horizontal trajectories, it is defined by
\begin{align}
(\mathrm{v}\cdot \hat{A})\Psi_0[A_{\bar{z}}]&=\left( v \frac{\pi}{k}\frac{\delta}{\delta A_{\bar{z}}}+\bar{v}A_{\bar{z}} \right)\Psi_0[A_{\bar{z}}]\equiv 0\\
\Rightarrow \Psi_0[A_{\bar{z}}]&=C\exp\left(-\frac{k}{2\pi}\int_\Sigma d^2x \frac{\bar{v}}{v}A_{\bar{z}}^2 \right).\label{Psi0higherg}
\end{align}
As for genus one, here the constant $C$ may depend on the complex structure and is fixed by properly normalizing the 
vacuum partition function. 
We show next that integrating it over the gauge orbit does result in a particular vector in the Hilbert space
\eqref{ZZKahler} obtained from K\"ahler quantization, holomorphic in the complex structure $\Omega$. 
In particular, we will see that, after integration, the vector $\mathrm{v}$  appears only in the Weyl anomaly. 

Given any seed wave function $\Psi_0$, the gauge integral is given by a  generalization of \eqref{UPsi0torus} on the torus. 
It reads
\begin{align}
\Psi[A_{\bar{z}}]&\equiv \int D'\lambda (U(\lambda)\cdot \Psi_0)[A_{\bar{z}}]\label{UPsi0}\\
&=\int D'\lambda \exp{\left(- \frac{k}{2\pi}\int_\Sigma d^2x \partial_z \lambda \partial_{\bar{z}}\lambda+\frac{k}{\pi}\int_\Sigma d^2x \partial_z \lambda A_{\bar{z}} \right)}\Psi_0[{}^\lambda A_{\bar{z}}]\label{Psihigherg},
\end{align}
where $A_{\bar{z}}\longrightarrow {}^\lambda  A_{\bar{z}}\equiv A_{\bar{z}}-\partial_{\bar{z}}\lambda$. 

We begin by decomposing 
\begin{subequations}
\begin{align}
A_{\bar{z}}&=\partial_{\bar{z}}\chi+\partial_{\bar{z}}\chi'\equiv \partial_{\bar{z}}\chi+i\pi u (\Im\Omega)^{-1}\overline{\omega}(\bar{z}),\;\overline{\omega}\equiv\overline{\omega}(\bar{z})d\bar{z},\\
\lambda&=\lambda_0+\lambda',\\
\partial_{\bar{z}}\lambda'&= i\pi (m+n\Omega)(\Im\Omega)^{-1}\overline{\omega}(\bar{z}),\; \partial_z \lambda'= -i\pi (m+n\overline{\Omega})(\Im\Omega)^{-1} \omega(z),\\
\mathcal{U}\overline{\omega}(\bar{z})&\equiv \partial_{\bar{z}}\chi'-\partial_{\bar{z}}\lambda'=i\pi(u-(m+n\Omega))(\Im\Omega)^{-1}\overline{\omega}(\bar{z}),\;m,n\in\mathbb{Z}^g,
\end{align}\label{chichiprime}
\end{subequations}
where both $\chi$, $\lambda_0$ are single-valued on $\Sigma$. The multivalued function $\lambda'$ appears everywhere only
through its derivatives, which are also single-valued on $\Sigma$.  Next we evaluate the integrand:
\begin{align}
&\quad S\equiv \ln \left(U(\lambda)\cdot \Psi_0\right)-\log{C} \nonumber\\
&=-\frac{k}{2\pi}\int_\Sigma d^2x \partial_z \lambda \partial_{\bar{z}}\lambda+\frac{k}{\pi}\int_\Sigma d^2x \partial_z \lambda A_{\bar{z}}-\frac{k}{2\pi}\int_\Sigma d^2x \frac{\bar{v}}{v}(A_{\bar{z}}-\partial_{\bar{z}}\lambda)^2\\
&=-\frac{k}{2\pi}\int_\Sigma d^2x \frac{1}{v} \partial_h \lambda_0 \partial_{\bar{z}}\lambda_0+\frac{k}{\pi}\int_\Sigma d^2x \frac{1}{v} \partial_h \lambda_0 \partial_{\bar{z}} \chi+\frac{k}{\pi}\int_\Sigma d^2x \frac{1}{v} \partial_h \lambda_0 \mathcal{U}\overline{\omega}(\bar{z})\label{vAS01} \nonumber \\
&\quad  -\frac{k}{2\pi}\int_\Sigma d^2x \partial_z \lambda' \partial_{\bar{z}}\lambda'+\frac{k}{\pi}\int_\Sigma d^2x \partial_z \lambda' \partial_{\bar{z}}\chi'
\nonumber\\
&\quad -\frac{k}{2\pi}\int_\Sigma d^2x \frac{\bar{v}}{v}\left[ \partial_{\bar{z}} \lambda' \partial_{\bar{z}}\lambda'-2\partial_{\bar{z}} \lambda' (\partial_{\bar{z}}\chi+\partial_{\bar{z}}\chi')+\partial_{\bar{z}}\chi \partial_{\bar{z}}\chi+2\partial_{\bar{z}}\chi \partial_{\bar{z}}\chi'+\partial_{\bar{z}}\chi' \partial_{\bar{z}}\chi'\right]. 
\end{align}\label{action-m2}
(On the torus, this reduces to \eqref{torusgaugeint} with $a_1(0)=0$). Let us look now at the terms involving $\lambda_0$:
\begin{align}
\tilde{S}&\equiv -\frac{k}{2\pi}\int_\Sigma d^2x \frac{1}{v} \partial_h \lambda_0 \partial_{\bar{z}}\lambda_0+\frac{k}{\pi}\int_\Sigma d^2x \frac{1}{v} \partial_h \lambda_0 \partial_{\bar{z}} \chi+\frac{k}{\pi}\int_\Sigma d^2x \frac{1}{v} \partial_h \lambda_0 \mathcal{U}\overline{\omega}(\bar{z}).\label{Stilde0}
\end{align}
Since $\tilde{S}$ is quadratic, the $\lambda_0$-integral in \eqref{Psihigherg} can be evaluated exactly. The saddle point $\lambda_{0,\text{cl}}$ at which $\tilde{S}$ is extremal satisfies the equation of motion
\begin{align}
\partial_{\bar{z}}\left[ \partial_h(\lambda_{0,\text{cl}}-\chi)-\bar{v}\mathcal{U}\overline{\omega}(\bar{z}) \right]=0.\label{lambda0eom1}
\end{align}
In addition, we recall that any horizontal trajectory $\gamma(s)\sim\sum_I N_I a_I$ is homologous to a linear combination of the $a$-cycles and that the trajectory is an integral curve of the vector $v$, so that $v(z(s))=\frac{dz}{ds}$. We thus have 
$\oint_\gamma\partial_h (\lambda_{0,\text{cl}}-\chi)=0$, since $\lambda_{0,\text{cl}}$ and $\chi$ are periodic. Then \eqref{lambda0eom1} is integrated to
\begin{align}
\partial_h(\lambda_{0,\text{cl}}-\chi)=-v\mathcal{U}\omega(z)+\bar{v}\mathcal{U}\overline{\omega}(\bar{z}).\label{lambda0eom2}
\end{align}
Substituting \eqref{lambda0eom2} back into the action \eqref{Stilde0} and including the fluctuation $f$ around the saddle point 
$\lambda_{0,\text{cl}}$, the action $S$ becomes
\begin{align}
S&=-\frac{k}{2\pi}\int_\Sigma d^2x \frac{1}{v} \partial_h f\partial_{\bar{z}}f+\frac{k}{2\pi}\int_\Sigma d^2x \partial_z \chi \partial_{\bar{z}}\chi\\
&\quad -\frac{k}{2\pi}\int_\Sigma d^2x \mathcal{U} \omega(z)\mathcal{U}\overline{\omega}(\bar{z})-\frac{k}{2\pi}\int_\Sigma d^2x \partial_z \lambda' \partial_{\bar{z}}\lambda'+\frac{k}{\pi}\int_\Sigma d^2x \partial_z \lambda' \partial_{\bar{z}}\chi' \nonumber.
\end{align}
Thus, we see that the vector field $\mathrm{v}$ indeed drops out, except in the fluctuation term. 

Substituting the
definitions~\eqref{chichiprime} into $\Psi[A_{\bar{z}}]$ and repeating the same calculation done in the torus case, we arrive at 
\begin{align}
\Psi[A_{\bar{z}}]&\equiv \int D'\lambda (U(\lambda)\cdot \Psi_0)[A_{\bar{z}}]\\
&=C \det{}^{-\frac{1}{2}}\left(-{\frac{k}{2\pi}}\partial_{\bar{z}}\partial_h\right)e^{+\frac{k\pi}{2}u(\Im \Omega)^{-1} u+\frac{k}{2\pi} \int_\Sigma d^2x \partial_z \chi \partial_{\bar{z}}\chi}\\
&\qquad \qquad \times  \sum_{m,n\in\mathbb{Z}^g}\exp\left(-2\pi i kun+i\pi k(m+n\Omega)n\right)\nonumber\\
&= e^{+\frac{k}{2\pi} \int_\Sigma d^2x \partial_z \chi \partial_{\bar{z}}\chi}
\frac{e^{+\frac{k\pi}{2}u(\Im \Omega)^{-1} u}}{\widetilde{F}(\Omega)^{\frac{1}{2}}}\theta\begin{bmatrix}0\\0\end{bmatrix}(ku,k\Omega),
\label{genus-g-m3}
\end{align}
which is \eqref{ZZKahler} with $\mu=0$.  When computing the quadratic functional integral  over the fluctuations $f$ 
(with the zero modes discarded), we used again the freedom in choosing the constant $C$ to ensure that the wave function is a 
section of a projectively flat bundle over the complex structure moduli space. For Abelian Chern-Simons theories, this is
achieved by using the factorization of the Laplacian~\eqref{ztl} (or by comparison with the formulas in~\cite{Bos:1989kn}), 
which results in
$C\det^{-1/2}(-{\frac{k}{2\pi}}\partial_{\bar{z}}\partial_h)=1/\widetilde{F}(\Omega)^{\frac{1}{2}}$. Note that the constant $C$ also
reabsorbs a term that contains a Weyl anomaly and therefore, because of \eqref{vmetric}, a dependence on $\mathrm{v}$. 
We also used $k/2\in\mathbb{Z}$ and $m,n\in\mathbb{Z}^g$, and discarded the trivial sum 
over $m$, that is the sum over large gauge transformations that can be extended to the bulk and under which the 
wave function is invariant.

\subsubsection{Wilson loops}
On a higher-genus handlebody, besides Wilson loops that can be regarded as ``world histories of  mesons," there is also another 
class of gauge-invariant observables, which correspond to the ``world histories of baryons" running along the non-contractible cycles; 
see 
\cite{Witten:1989wf}. For the  Abelian case that we have considered here, however, the fusion rule is trivial, so those ``baryon world histories" can be decomposed into disjoint Wilson loops running along the non-contractible cycles of the handlebody. Therefore, it 
suffices to consider only standard Wilson loops.

We would like to generalize the ``blowing-up" of Wilson loops that we studied on the torus in \hyperref[WLtorusPsi]{Section \ref*{WLtorusPsi}} to higher genus. Consider the loops $C_I$ running along the $g$ non-contractible cycles of $M$ and endowed
 with charges $\mu_I\in\mathbb{Z}_k$. The resulting Wilson loops are then 
 ``blown up" into  operators $\hat{W}[\Sigma,\mathrm{w}]$ on 
 $\Sigma$, parametrized by real one-forms
\begin{align}
\mathrm{w}&=(\underbrace{w_0 dz+w'\omega}_{\equiv wdz})+(\underbrace{\bar{w}_0d\bar{z}+\bar{w}'\overline{\omega}}_{\equiv \bar{w}d\bar{z}})\equiv \mathrm{w}_0+\mathrm{w}',
\end{align}
where the $w_0$ are periodic on $\Sigma$ and $w'_I$ are constant. A ``blown-up" Wilson loop operator is then
\begin{align}
\hat{W}[\Sigma,\mathrm{w}]&\equiv \exp\left( i\int_\Sigma d^2x (\bar{w}\hat{A}_z+w\hat{A}_{\bar{z}}) \right),\label{Www}
\end{align}
Importantly, gauge invariance of $\hat{W}[\Sigma,\mathrm{w}]$ demands that
\begin{subequations}
\begin{empheq}[left=\empheqlbrace]{align}
&\partial_z \bar{w}_0+ \partial_{\bar{z}} w_0=0\quad\Rightarrow\quad w_0=i\partial_z \eta  , \\
&\int_\Sigma d^2x (\bar{w}' \overline{\omega}(\bar{z}) \partial_z \lambda'+w' \omega(z) \partial_{\bar{z}} \lambda') \in 2\pi \mathbb{Z}.\end{empheq}\label{w0eta}
\end{subequations}
Here $\eta$ is a real single-valued function on $\Sigma$, and $\lambda'$ is a large gauge transformation \eqref{chichiprime}.
As discussed in \hyperref[WLtorusPsi]{Section \ref*{WLtorusPsi}}, by demanding large gauge invariance on the ``blown-up" surface
 one selects a class of preferred framings, satisfying
\begin{align}
-i\pi m(\bar{w}'-w')+i\pi n(\Omega w'-\overline{\Omega}\bar{w}')\in 2\pi\mathbb{Z}.
\end{align}
Inverting this condition yields
\begin{align}
\bar{w}'&=(\Im\Omega)^{-1}(\Omega\mu+N),\;\mu,N\in \mathbb{Z}^g.
\end{align}
For $\eta=0$ on the torus, we recover the Wilson loop $\hat{W}[\Sigma;\mu,N]$ in \eqref{torusWmuN}.

The gauge invariance of $\hat{W}[\Sigma,\mathrm{w}]$ allows us to commute it with the gauge integral \eqref{UPsi0}, and let it
act on the
 seed wave function $\Psi_0[A_{\bar{z}}]$. As seed wave function we use the wave function of trivial holonomy along the 
 contractible cycles given in \eqref{Psi0higherg}:
\begin{align}
\Psi_0[A_{\bar{z}}]&=C \exp\left(-\frac{k}{2\pi}\int_\Sigma d^2x \frac{\bar{v}}{v}A_{\bar{z}}^2 \right).
\end{align}
Using again the Baker-Campbell-Hausdorff formula \eqref{BCH}, one gets
\begin{align}
&\quad \Psi_{\mathrm{w},0}[A_{\bar{z}}]\equiv \hat{W}[\Sigma,\mathrm{w}]\Psi_0[A_{\bar{z}}]\\
&=C \exp\left( -\frac{k}{2\pi}\int_\Sigma d^2x \frac{\bar{v}}{v}A_{\bar{z}}^2 -i\int_\Sigma d^2x \left[\bar{w}\left(\frac{\bar{v}}{v}\right)-w\right] A_{\bar{z}}+\frac{\pi}{2k} \int_\Sigma d^2x \bar{w} \left[\bar{w}\left(\frac{\bar{v}}{v}\right)-w\right] \right).
\end{align}
We repeat the calculation done in the last subsection to evaluate the gauge integral \eqref{UPsi0}. The integrand is
\begin{align}
S&\equiv \ln \left(U(\lambda)\cdot \Psi_{\mathrm{w},0}\right) -\log(C)\\
&=-\frac{k}{2\pi}\int_\Sigma d^2x \partial_z \lambda \partial_{\bar{z}}\lambda+\frac{k}{\pi}\int_\Sigma d^2x \partial_z \lambda A_{\bar{z}}-\frac{k}{2\pi}\int_\Sigma d^2x \frac{\bar{v}}{v}(A_{\bar{z}}-\partial_{\bar{z}}\lambda)^2\\
&\quad -i\int_\Sigma d^2x \left[\bar{w}\left(\frac{\bar{v}}{v}\right)-w\right] (A_{\bar{z}}-\partial_{\bar{z}}\lambda)+\frac{\pi}{2k} \int_\Sigma d^2x \bar{w} \left[\bar{w}\left(\frac{\bar{v}}{v}\right)-w\right]  \nonumber\\
&=-\frac{k}{2\pi}\int_\Sigma d^2x \frac{1}{v} \partial_h \lambda_0 \partial_{\bar{z}}\lambda_0+\frac{k}{\pi}\int_\Sigma d^2x \frac{1}{v} \partial_h \lambda_0 (\partial_{\bar{z}} \chi+\mathcal{U}\overline{\omega}(\bar{z}))\label{vAS02}\\
&\quad  +i\int_\Sigma d^2x \partial_{\bar{z}} \lambda_0 \left[\bar{w}\left(\frac{\bar{v}}{v}\right)-w\right]-\frac{k}{2\pi}\int_\Sigma d^2x \partial_z \lambda' \partial_{\bar{z}}\lambda'+\frac{k}{\pi}\int_\Sigma d^2x \partial_z \lambda' \partial_{\bar{z}}\chi'
\nonumber\\
&\quad +i \int_\Sigma d^2x  \partial_{\bar{z}}\lambda' \left[\bar{w}\left(\frac{\bar{v}}{v}\right)-w\right]-i \int_\Sigma d^2x \left[\bar{w}\left(\frac{\bar{v}}{v}\right)-w\right] A_{\bar{z}} \nonumber\\
&\quad + \frac{\pi}{2k} \int_\Sigma d^2x \bar{w} \left[\bar{w}\left(\frac{\bar{v}}{v}\right)-w\right] \nonumber\\
&\quad -\frac{k}{2\pi}\int_\Sigma d^2x \frac{\bar{v}}{v}\left[ \partial_{\bar{z}}\chi \partial_{\bar{z}}\chi+2\partial_{\bar{z}}\chi \partial_{\bar{z}}\chi'+ \mathcal{U}\omega(z)\mathcal{U}\overline{\omega}(\bar{z})  -2\partial_{\bar{z}} \lambda' \partial_{\bar{z}}\chi\right] .\nonumber
\end{align}
The terms containing $\lambda_0$ are
\begin{align}
\tilde{S}&\equiv -\frac{k}{2\pi}\int_\Sigma d^2x \frac{1}{v} \partial_h \lambda_0 \partial_{\bar{z}}\lambda_0+\frac{k}{\pi}\int_\Sigma d^2x \partial_z \lambda_0 \partial_{\bar{z}} \chi\\
&\qquad\qquad +\frac{k}{\pi}\int_\Sigma d^2x  \partial_{\bar{z}} \lambda_0 \left(\frac{\bar{v}}{v} \partial_{\bar{z}}\chi+ \frac{\bar{v}}{v}\mathcal{U}\overline{\omega}(\bar{z})-\frac{i\pi}{k}w+ \frac{i\pi}{k} \frac{\bar{v}}{v} \bar{w} \right)\nonumber\\
&= -\frac{k}{2\pi}\int_\Sigma d^2x \frac{1}{v} \partial_h \lambda_0 \partial_{\bar{z}}\lambda_0+\frac{k}{\pi}\int_\Sigma d^2x \partial_z \lambda_0 \partial_{\bar{z}}( \chi +\frac{\pi}{k}\eta )\label{Stilde1}\\
&\qquad\qquad +\frac{k}{\pi}\int_\Sigma d^2x \frac{\bar{v}}{v} \partial_{\bar{z}} \lambda_0 \left( \partial_{\bar{z}}(\chi+ \frac{\pi}{k}\eta)+(\mathcal{U}+\frac{i\pi}{k}\bar{w}')\overline{\omega}(\bar{z}) \right) \nonumber.
\end{align}
The saddle point $\lambda_{0,\text{cl}}$ at which $\tilde{S}$ is extremal satisfies the equation of motion
\begin{align}
\partial_{\bar{z}}\left[ \partial_h(\lambda_{0,\text{cl}}-\chi- \frac{\pi}{k}\eta)-\bar{v}(\mathcal{U}+\frac{i\pi}{k}\bar{w}')\overline{\omega}(\bar{z}) \right]=0,\label{lambda0eom3}
\end{align}
which is integrated to
\begin{align}
\partial_h\lambda_{0,\text{cl}}=\partial_h (\chi+ \frac{\pi}{k}\eta)  -v(\mathcal{U}+\frac{i\pi}{k}\bar{w}')\omega(z)+\bar{v}(\mathcal{U}+\frac{i\pi}{k}\bar{w}')\overline{\omega}(\bar{z}).\label{lambda0eom4}
\end{align}
Substituting \eqref{lambda0eom4} back to the action \eqref{Stilde1} and including the fluctuation $f$ around the saddle point, we get
\begin{align}
S&=-\frac{k}{2\pi}\int_\Sigma d^2x \frac{1}{v} \partial_h f\partial_{\bar{z}}f+\frac{k}{2\pi}\int_\Sigma d^2x \partial_z \chi \partial_{\bar{z}}\chi\\
&\quad -\frac{k}{2\pi}\int_\Sigma d^2x \mathcal{U} \omega(z)\mathcal{U}\overline{\omega}(\bar{z})-\frac{k}{2\pi}\int_\Sigma d^2x \partial_z \lambda' \partial_{\bar{z}}\lambda'+\frac{k}{\pi}\int_\Sigma d^2x \partial_z \lambda' \partial_{\bar{z}}\chi' \nonumber\\
&\quad  -i \int_\Sigma d^2x \mathcal{U}\overline{\omega}(\bar{z}) (\bar{w}'-w') \omega(z) +\frac{\pi}{2k} \int_\Sigma d^2x (\bar{w}'-w') \omega(z) \bar{w}' \overline{\omega}(\bar{z})\nonumber.
\end{align}
So, once again, $v$ drops out of the action--- except in the fluctuation term which also gives the Weyl anomaly. Moreover, the function 
$\eta$ in $\mathrm{w}$ drops out as well. Evaluating the path integral in the same way as before, we finally get
\begin{align}
\Psi_{\mathrm{w}}[A_{\bar{z}}]&\equiv \int D'\lambda (U(\lambda)\cdot\Psi_{\mathrm{w},0})[A_{\bar{z}}]\\
&= C\det{}^{-1/2}\left(-{\frac{k}{2\pi}} \partial_{\bar{z}} \partial_h\right) e^{+\frac{k\pi}{2}u(\Im \Omega)^{-1} u +\frac{k}{2\pi} \int_\Sigma d^2x \partial_z \chi \partial_{\bar{z}}\chi} \\
&\qquad\qquad \times \sum_{m,n\in\mathbb{Z}^g}\exp\left( -2\pi i kun+i\pi k(m+n\Omega)n  \right) \nonumber\\
&\qquad\qquad \times \exp\left(  +2\pi i u\mu-2\pi i m\mu -i\pi n\Omega \mu +\frac{\pi}{2k}(2i\mu)(\Omega \mu+N) \right) \nonumber\\
&=e^{+\frac{k}{2\pi} \int_\Sigma d^2x \partial_z \chi \partial_{\bar{z}}\chi}
\frac{e^{+\frac{k\pi}{2}u(\Im \Omega)^{-1} u}}{\widetilde{F}(\Omega)^{\frac{1}{2}}}
\theta\begin{bmatrix} \frac{\mu}{k} \\0\end{bmatrix}(ku,k\Omega)e^{\frac{i\pi}{k}\mu N}.
\end{align}
Here used the same normalization for $C$ as in eq.~\eqref{genus-g-m3} together with $k/2\in\mathbb{Z}$, $m,n\in\mathbb{Z}^g$ 
and $\mu\in\mathbb{Z}$, and discarded the trivial sum over $m$. 

Since $\eta$ drops out eventually, we can repeat the analysis we performed in the torus case. Namely, 
\begin{align}
\hat{W}[\Sigma,\mathrm{w}']&\equiv \exp\left( i\int_\Sigma d^2x (\bar{w}'\hat{A}_z+w'\hat{A}_{\bar{z}}) \right),\;\bar{w}'=(\Im\Omega)^{-1}(\Omega\mu+N),\;\mu,N\in \mathbb{Z}^g\\
&=\exp\left[ i \mu \int_\Sigma d^2x \left( \Omega (\Im\Omega)^{-1} \overline{\omega}(\bar{z})  \hat{A}_z+ \overline{\Omega} (\Im\Omega)^{-1} \omega(z)  \hat{A}_{\bar{z}}\right)\right. \\
&\qquad\qquad \left. +i N (\Im\Omega)^{-1} \int_\Sigma d^2x \left(  \overline{\omega}(\bar{z})  \hat{A}_z+ \omega(z)  \hat{A}_{\bar{z}}\right) \right]\nonumber\\
&=\exp\left[ i \mu \int_\Sigma d^2x \left( \Omega (\Im\Omega)^{-1} \overline{\omega}(\bar{z})  \hat{A}_z+ \overline{\Omega} (\Im\Omega)^{-1} \omega(z)  \hat{A}_{\bar{z}}\right)\right]\\
&\quad \times\exp\left[ +i N (\Im\Omega)^{-1} \int_\Sigma d^2x \left(  \overline{\omega}(\bar{z})  \hat{A}_z+ \omega(z)  \hat{A}_{\bar{z}}\right) \right] \exp \left( +\frac{i\pi}{k}\mu N \right) \nonumber\\
&\equiv \hat{W}[\Sigma,\mu]\hat{W}[\Sigma,N]e^{ +\frac{i\pi}{k}\mu N }.
\end{align}
In the end, the discussion in the torus case generalizes to higher-genus cases. Namely, we can interpret $\hat{W}[\Sigma,\mu]$ and 
$\hat{W}[\Sigma,N]$ respectively as the ``blowing-up" of loops along the non-contractible and contractible cycles of $M$. The only
 contribution of $\hat{W}[\Sigma,N]$ to a gauge-invariant wave function is the phase $\exp(+(i\pi/k)\mu N)$ and is due to the linking of 
the loops. In the special case $N_I=N'_{IJ}\mu_J$ for some symmetric matrix $N'$ with integer entries, this phase $\exp(+(i\pi/k)\mu N' \mu)$ is naturally interpreted as the framing anomaly.

\section{The non-Abelian Case}\label{secnonabel}
We consider now the non-Abelian case, with a compact, simply-connected and simple Lie group $G$ on a solid torus. In this
section $M$ is always the torus handlebody and $\Sigma=T^2$. 
By generalizing the
equation \eqref{cohalAphi} found in the Abelian case, we will consider the $A_\phi$-eigenstate $| A_\phi \rangle\rangle$ translated by
 the conjugate momentum $A_t$. In the coherent state basis, it reads
\begin{align}
(A_{\bar{z}}| A_\phi \rangle\rangle&=C\exp{\left( +\frac{k}{2\pi}\int_\Sigma d^2x\Tr A_{\bar{z}}^2-\frac{k}{\pi}\int_\Sigma d^2x \Tr A_{\bar{z}} A_\phi -\frac{ik }{4\pi}\tau\tau_2^{-1} \int_\Sigma d^2x \Tr A_\phi^2 \right)}.\label{cohalAphinAb}
\end{align}
The amplitude with $A_\phi|_{\Sigma_0}=a_\phi(0)$ constant is
\begin{align}
&\quad (A_{\bar{z}} |e^{-iHR}|a_\phi(0)\rangle\rangle\equiv \int_{A_\phi|_{\Sigma_0}=a_\phi(0)}DA_1 DA_2 \delta(F_{12}) e^{iI}\\
&=C\int_{A_\phi|_{\Sigma_0}=a_\phi(0)}DA_1 DA_2 \delta(F_{12})\exp{\left(+\frac{k}{2\pi}\int_{\Sigma_R}d^2x \Tr A_{\bar{z}}^2-\frac{k}{\pi}\int_{\Sigma_R}d^2x\Tr A_{\bar{z}} A_{\phi}(R) \right)}\label{nAbI}\\
&\qquad\times \exp \left(-\frac{ik}{4\pi }\tau\tau_2^{-1}\int_{\Sigma_R}d^2x\Tr (A_{\phi}(R))^2\right) \exp \left(+\frac{ik}{2\pi}\int_M d^2x dr \Tr A_t\partial_r A_\phi \right).\nonumber
\end{align}
Solving the constraint $F_{12}=0$ by \cite{Elitzur:1989nr}
\begin{align}
A_i&=g^{-1}a_ig+g^{-1}\partial_i g,\;a_i:[0,R]\rightarrow G,\;g:[0,R]\times \Sigma\rightarrow G,\;i=\phi,t,
\end{align}
the action \eqref{nAbI} becomes
\begin{align}
iI&=\frac{k}{2\pi}\int_{\Sigma_R} d^2x \Tr A_{\bar{z}}^2-\frac{ik}{12\pi}\int_M d^2x dr \Tr[(g^{-1}dg)^3]-\frac{ik}{4\pi }\tau\tau_2^{-1}\int_{\Sigma_R} d^2x \Tr(a_\phi(R))^2\nonumber\\
&\quad +\frac{k}{2\pi}\int_{\Sigma_R} d^2x \Tr[g \partial_\phi ((\partial_{\bar{z}}+2A_{\bar{z}})g^{-1})]-2\frac{k}{2\pi}\int_{\Sigma_R} d^2x \Tr[a_{\phi}(R) g (\partial_{\bar{z}}+A_{\bar{z}}) g^{-1}]\nonumber\\
&\quad +\frac{ik}{2\pi}\int_M d^2x dr \Tr[-(a_\phi a_t-a_t a_\phi )\partial_r gg^{-1}+a_t \partial_r a_\phi]+\log(C)  \label{nAbI2}.
\end{align}
On the torus $a_\phi(r)$ and $a_t (r)$  commute so they are elements of the Cartan subalgebra $\fr{h}$ of $\fr{g}$; this is not true in general for higher genus. Moreover, by integrating out $a_t(r)$ we get $a_\phi(r)=a_\phi(0)$, so the amplitude \eqref{nAbI} becomes
\begin{align}
&\quad (A_{\bar{z}} |e^{-iHR}|a_\phi(0)\rangle\rangle= Z(A_{\bar{z}},a_\phi(0);\tau)=\int Dg e^{iI},\;\label{cWZWnAb}\\
iI&=-\frac{ik}{12\pi}\int_M d^2x dr \Tr[(g^{-1}dg)^3]+\frac{k}{2\pi}\int_{\Sigma_R} d^2x \Tr A_{\bar{z}}^2\\
&\quad -\frac{ik}{4\pi}\tau\tau_2^{-1}\int_{\Sigma_R} d^2x \Tr(a_\phi(R))^2 +\frac{k}{2\pi}\int_{\Sigma_R} d^2x \Tr[g \partial_\phi ((\partial_{\bar{z}}+2A_{\bar{z}})g^{-1})] \nonumber\\
&\quad  -2\frac{k}{2\pi}\int_{\Sigma_R} d^2x \Tr[a_{\phi}(R) g (\partial_{\bar{z}}+A_{\bar{z}}) g^{-1}] +\log(C)\nonumber.
\end{align}
With an appropriate, $A_{\bar{z}}$-independent choice of $C$, this is the chiral Wess-Zumino-Witten path integral. For $a_\phi(0)=2\pi \mu/k$ where $\mu$ is an integral weight of $G$ and $A_{\bar{z}}=iu \tau_2^{-1}$, ref.~\cite{Perret:1990bc} shows that the path integral gives the Weyl-Kac character $\chi_{\mu,k}(u,\tau)$:
\begin{align}
(A_{\bar{z}} |e^{-iHR}|2\pi\mu/k\rangle\rangle&=e^{-\frac{k\pi}{2}\Tr[u\tau_2^{-1}u]} \chi_{\mu,k}(u,\tau)=e^{-\frac{k\pi}{2}\Tr[u\tau_2^{-1}u]}\frac{\theta^-_{\mu+\rho,k+h^\vee}(u,\tau)}{\theta^-_{\rho,h^\vee}(u,\tau)},\label{perretWK}
\end{align}
where $\rho$ and $h^\vee$ are respectively the Weyl vector and the dual Coxeter number of $\fr{g}$. The Weyl-odd theta function is defined as
\begin{align}
\theta^-_{\mu,k}(u,\tau)&\equiv \sum_{w\in W}\epsilon(w)\theta_{\mu,k}(w(u),\tau)
\end{align}
where $W$ is the Weyl group of $G$ and $\epsilon(w)$ is the signature of $w\in W$. $\theta_{\mu,k}(u,\tau)$ is the level-$k$ theta function for the Lie algebra $\fr{g}$, whose definition is recalled in \eqref{gsimpletheta1}.

\subsection{Wilson loops}
The Wilson loop operator of the representation generated by the integral highest weight $\mu$ of $G$, along a loop $C$ of constant radius in $M$, is
\begin{align}
\hat{W}_{\mu}[C]&\equiv \Tr_{\mu}\mathcal{P}\exp{\left( \oint_C \hat{A} \right)}=\Tr_{\mu}\mathcal{P}\exp{\left( \oint_C \hat{a} \right)}.
\end{align}
In the last equality, we stripped off the pure gauge part of $\hat{A}$ (recall the definition $A_i=g^{-1}a_i g+g^{-1}\partial_i g$) due to the trace in the definition of $\hat{W}_{\mu}[C]$, so we only need to look at the equal-radius canonical commutation relation of $\hat{a}_i(r)$,
which we read off from \eqref{nAbI2}:  
\begin{align}
[\hat{a}_\phi^{\fr{j}}(r),\hat{a}_t^{\fr{l}}(r)]&=-\frac{2\pi}{k}\delta^{\fr{j}\fr{l}}.
\end{align}
Here we have expanded $a_{\phi,t}(r)=\sum_{\fr{j}=1}^{\text{rank}(\fr{g})}a^{\fr{j}}_{\phi,t}(r)H^{\fr{j}}$ in the Cartan-Weyl basis $\{H^{\fr{j}}\}$ of the Cartan subalgebra $\fr{h}$ of $\fr{g}$, where
 ${\fr{j}},{\fr{l}}=1,\ldots,\text{rank}(\fr{g})$. For the loop $C_t$ at $r=0$ running along the $t$-direction, 
\begin{align}
\hat{W}_{\mu}[C_t]&=\Tr_{\mu}\mathcal{P}\exp{\left( \oint_{C_t} \hat{a} \right)}.\label{nawl}
\end{align}
As in the Abelian case, we map \eqref{nawl} to a ``blown up'' gauge-invariant operator $\hat{W}_{\mu}[\Sigma]$ 
defined on $\Sigma$, which is to be identified with the translation operator by the conjugate momentum $a^{\fr{j}}_t$, acting on the 
$a_\phi=0$ eigenstate $|0\rangle$. Since $\hat{a}_t$ is constant on $\Sigma$, the Wilson loop is  simply given by
\begin{align}
\hat{W}_{\mu}[C_t]&\longrightarrow \hat{W}_\mu[\Sigma]\equiv \Tr_{\mu}\exp{\left(\hat{a}_t \right)}=\chi_\mu(\hat{a}_t),
\end{align}
The first equality is the character of $\hat{a}_t$ as an element of $\fr{h}$. This is expressed as a Weyl character in the second equality. We recall the latter's definition:
\begin{align}
\chi_\mu(\hat{a}_t)&\equiv \sum_{\mu'\in\Omega_\mu}\exp{\left[(\mu',\hat{a}_t) \right]},\label{weylch1}
\end{align}
where $\mu'$ are the weights in the weight system $\Omega_\mu$ of the highest weight $\mu$, which span a highest-weight representation of $G$. By the Weyl character formula, \eqref{weylch1} can be written as a ratio of sums over the Weyl group $W$ of $G$:
\begin{align}
\chi_\mu(\hat{a}_t)&=\frac{D_{\mu+\rho}(\hat{a}_t)}{D_\rho (\hat{a}_t)},\text{ where }D_{\mu}(\hat{a}_t)\equiv \sum_{w\in W}\epsilon(w)e^{(w(\mu),\hat{a}_t)}.\label{weylchformula}
\end{align}

Now, we act with $\hat{W}_\mu[\Sigma]$ on the $a_\phi=0$ eigenstate $|0\rangle$, radially evolve it and compute the 
overlap with the coherent state $(A_{\bar{z}}|$ with a constant field as final condition: $A_{\bar{z}}=iu \tau_2^{-1}$. By using
\eqref{perretWK} we get
\begin{align}
&\quad (A_{\bar{z}}=iu \tau_2^{-1} |e^{-iHR}\hat{W}_\mu[\Sigma]|0\rangle\\
&=\sum_{\mu'\in\Omega_\mu} (A_{\bar{z}}=iu \tau_2^{-1} |e^{-iHR}|2\pi \mu'/k\rangle\rangle\\
&=e^{-\frac{k\pi}{2}\Tr[u\tau_2^{-1}u]}\sum_{\mu'\in\Omega_\mu} \chi_{\mu',k}(u,\tau) .\label{amplitudeexpect}
\end{align}
The Wilson loop operator $\hat{W}_{\mu}[C_t]$ should compute the Weyl-Kac character when inserted into the path integral --as expected from a canonical quantization. Since $\hat{W}_\mu[\Sigma]$ was obtained by ``blowing-up'' $\hat{W}_{\mu}[C_t]$, we expect the amplitude $(A_{\bar{z}} |e^{-iHR}\hat{W}_\mu[\Sigma]|0\rangle$ to give the same result, namely Eq.~\eqref{amplitudeexpect}. For this to be valid, the Weyl-Kac character needs to satisfy the identity
\begin{align}
\sum_{\mu'\in\Omega_\mu} \chi_{\mu',k}(u,\tau) = \chi_{\mu,k}(u,\tau).
\end{align}
Because the radial evolution is linear in the initial state, this identity holds if the corresponding identity is true for the Weyl character of the Lie algebra,
\begin{align}
\sum_{\mu'\in\Omega_\mu} \chi_{\mu'}&=\chi_\mu.\label{mysteriousidentity}
\end{align}
This should be understood as an equality in terms of the Weyl character formula \eqref{weylchformula}. Intuitively, this identity should hold due to the fact that all the weights $(\mu'+\rho )$ with $\mu'\in\Omega_\mu$, except for the highest weight $\mu$, pair up 
 under simple Weyl transformations. 

As an example, let us look at the $G=SU(2)$ case, in the spin-$J$ representation where $2J\in\mathbb{Z}_{\geq 0}$. 
In this case we have
\begin{align}
\hat{W}_J[\Sigma] |0\rangle&\equiv \Tr_{J}\exp{\left(\hat{a}_t \right)}|0\rangle=\sum_{m=-J}^J \exp{\left(-im \hat{a}_t \right)}|0\rangle=\sum_{m=-J}^J  |2\pi m/k\rangle\rangle.
\end{align}
The amplitude is
\begin{align}
 (A_{\bar{z}}=iu \tau_2^{-1} |e^{-iHR}\hat{W}_J[\Sigma]|0\rangle&=
 e^{-\frac{k\pi}{2}\Tr[u\tau_2^{-1}u]}\sum_{m=-J}^J \chi_{m,k}(u,\tau), \label{amplitudeexpectsu2}
\end{align}
where here we used Eq.~\eqref{perretWK}.
So the question is whether the following identity holds
\begin{align}
\sum_{m=-J}^J \chi_{m,k}(u,\tau)= \chi_{J,k}(u,\tau)\label{mysteriousidentitysu2} .
\end{align}
To prove this, note that (the numerator of) the $\fr{su}(2)$ Weyl character trivially satisfies \eqref{mysteriousidentity}:
\begin{align}
\sum_{m=-J}^J \sin((2m+1)2\pi u)=\sin((2J+1)2\pi u),\label{weylsu2trivial}
\end{align}
The proof that this implies \eqref{mysteriousidentitysu2} is achieved by first writing down explicitly the Weyl-Kac character
\begin{align}
&\chi_{J,k}(u,\tau)\nonumber\\
&=\frac{\sum_{n\in\mathbb{Z}}q^{\frac{1}{k+2}(J+1/2+n(k+2))^2-1/4)}\left( e^{4\pi i u (J+1/2+n(k+2))}-e^{-4\pi i u (J+1/2+n(k+2))} \right)}{\left(e^{2\pi i u}-e^{-2\pi i u}\right)\prod_{l=1}^\infty (1-q^l)(1-q^l e^{4\pi i u} )(1-q^l e^{-4\pi i u})}\\
&=\frac{q^{-\frac{1}{4}}}{\sin(2\pi u)\prod_{l=1}^\infty (1-q^l)(1-q^l e^{4\pi i u} )(1-q^l e^{-4\pi i u})}\label{WKsu2explicit}\\
&\quad \times\sum_{n\in\mathbb{Z}}q^{\frac{1}{4(k+2)}(2J+1+2n(k+2))^2} \sin{\left( (2J+1+2n(k+2))2\pi u \right)},\nonumber
\end{align}
substituting it into \eqref{mysteriousidentitysu2}, and comparing both sides for each $|n|\in\mathbb{Z}_{\geq 0}$. 

For the $n=0$ terms, note that, by differentiating \eqref{weylsu2trivial} $2l$-times with respect to $u$, we get
\begin{align}
\frac{(-1)^l}{(2\pi)^{2l}}&\frac{d^{2l}}{du^{2l}}\eqref{weylsu2trivial},\quad l=0,1,\ldots\quad \Rightarrow\nonumber\\
&\sum_{m=-J}^J (2m+1)^{2l} \sin((2m+1)2\pi u)=(2J+1)^{2l} \sin((2J+1)2\pi u).\label{weylsu2diff1}
\end{align}
This implies
\begin{align}
&\quad\sum_{m=-J}^J e^{\alpha (2m+1)^2} \sin((2m+1)2\pi u),\;\alpha\in\mathbb{C}\\
&=\sum_{m=-J}^J\sum_{l=0}^\infty \frac{1}{l!}(2m+1)^{2l}\sin((2m+1)2\pi u)\\
&\overset{\eqref{weylsu2diff1}}{=}e^{\alpha (2J+1)^2}\sin((2J+1)2\pi u).
\end{align}
Taking $\alpha=1/(4(k+2))$, we prove the equality of the $n=0$ terms in \eqref{mysteriousidentitysu2}.

Next, for the $|n|\geq 1$ terms, we note that, for any $b\in\mathbb{C}$,
\begin{align}
&\quad \sin((2m+1+2b)2\pi u)+\sin((2m+1-2b)2\pi u)\nonumber\\
&=4\sin((2m+1)2\pi u)\sin(\pi/4-b2\pi u)\sin(\pi/4+b2\pi u),
\end{align}
and so \eqref{weylsu2trivial} implies
\begin{align}
&\quad\sum_{m=-J}^J\left[\sin((2m+1+2b)2\pi u)+\sin((2m+1-2b)2\pi u)\right]\nonumber\\
&=4\sin((2J+1)2\pi u)\sin(\pi/4-b2\pi u)\sin(\pi/4+b2\pi u)\\
&=\sin((2J+1+2b)2\pi u)+\sin((2J+1-2b)2\pi u).
\end{align}
Differentiating this identity $2l$-times with respect to $u$ gives
\begin{align}
&\quad \sum_{m=-J}^J \left[(2m+1+2b)^{2l} \sin((2m+1+2b)2\pi u)\right.\label{weylsu2diff2}\\
&\qquad\qquad \left.+(2m+1-2b)^{2l} \sin((2m+1-2b)2\pi u)\right]\nonumber\\
&=(2J+1+2b)^{2l} \sin((2J+1+2b)2\pi u)+(2J+1-2b)^{2l} \sin((2J+1-2b)2\pi u),\nonumber
\end{align}
and thus for any $b,\alpha\in\mathbb{C}$,
\begin{align}
&\quad \sum_{m=-J}^J \left[ e^{\alpha (2m+1+2b)^2}\sin((2m+1+2b)2\pi u)+e^{\alpha (2m+1-2b)^2}\sin((2m+1-2b)2\pi u)\right]\\
&=e^{\alpha (2J+1+2b)^2}\sin((2J+1+2b)2\pi u)+e^{\alpha (2J+1-2b)^2}\sin((2J+1-2b)2\pi u)   \nonumber.
\end{align}
Taking $\alpha=1/(4(k+2))$ and $b=|n|(k+2)$, we prove the equality of the $|n|\geq 1$ terms in \eqref{mysteriousidentitysu2}.

\subsection{Partition function as a gauge-invariant wave function}
Here we proceed in the same way as in the Abelian case. A wave function transforms as
\begin{align}
A_{\bar{z}}\longrightarrow {}^h  A_{\bar{z}}&\equiv hA_{\bar{z}}h^{-1}-h^{-1}\partial_{\bar{z}}h^{-1},\;h:\widetilde{\Sigma}\rightarrow G\\
(U(h)\cdot \Psi)[A_{\bar{z}}]&\equiv  \exp{\left(+\frac{k}{2\pi}\int_\Sigma d^2x \Tr[h^{-1}\partial_z h h^{-1}\partial_{\bar{z}}h]-\frac{ik}{12\pi}\int_M d^3x \Tr[(h^{-1}dh)^3]\right.}\\
&\qquad\qquad \left.-\frac{k}{\pi}\int_\Sigma d^2x \Tr[h^{-1}\partial_z h A_{\bar{z}}] \right)\Psi[{}^h A_{\bar{z}}].\nonumber
\end{align}
Since we consider a simply-connected group $G$, the gauge group $\mathcal{G}$ is connected. Similarly, 
starting from a wave function 
$\Psi_0[A_{\bar{z}}]$ that is not gauge-invariant, we can construct a gauge-invariant wave function by integrating over the gauge 
group:
\begin{align}
\Psi[A_{\bar{z}}]&\equiv \int_{\mathcal{G}} D'h (U(h)\cdot \Psi_0)[A_{\bar{z}}].\label{UPsi0nAb}
\end{align}
Taking $\Psi_0[A_{\bar{z}}]=(A_{\bar{z}}|a_{\phi}\rangle\rangle$ as the seed wave function in \eqref{cohalAphinAb} and after a quick calculation, one recovers the chiral Wess-Zumino-Witten path integral \eqref{cWZWnAb}. In other words, radially evolving the wave function is equivalent to integrating over the gauge group, which results in a gauge-invariant wave function.

\subsection*{Acknowledgements}
M.P.\ is supported in part by NSF grant PHY-1915219.

\begin{appendix}
\section{The Riemann Theta Function}\label{apptheta}
We define the genus-$g$ Riemann theta function $\theta[\substack{a\\b}](z,\Omega)$ with characteristics $(a,b)$ as in 
Mumford~\cite{Mumford:1338263},
\begin{align}
\theta[\substack{a\\b}](z,\Omega)&\equiv \sum_{n\in\mathbb{Z}^g}\exp{\left[ \pi i (n+a)\Omega (n+a)+2\pi i (n+a) (z+b) \right]},\;z\in \mathbb{C}^g,\;a,b\in\mathbb{R}^g,\label{mumford2}
\end{align}
with the quasi-periodicity
\begin{align}
\begin{cases}
\theta[\substack{a\\b}](z+m,\Omega)&=\exp{\left( 2\pi i a m \right)}\;\theta[\substack{a\\b}](z,\Omega),\;m\in\mathbb{Z}^g,\\
\theta[\substack{a\\b}](z+\Omega n,\Omega)&=\exp{\left( -2\pi i b n \right)}\exp{\left( -\pi i n\Omega n-2\pi i n z \right)}\;\theta[\substack{a\\b}](z,\Omega),\;n\in\mathbb{Z}^g.
\end{cases}
\end{align}
Consider a compact, simple and simply-connected group $G$. We identify the root space and the co-root space using the inner product $(\cdot,\cdot)$, for which we use the standard normalization such that the longest roots are of square-length two. We recall 
also the definition of the level-$k$ theta function on the genus-one Riemann surface (the torus) given in \cite{Axelrod:1989xt},
\begin{align}
\theta_{\mu,k}(u,\tau)&\equiv \sum_{\beta^\vee\in \Lambda^{R}}\exp{\left(\pi ik\tau \left(\beta^\vee+\frac{\mu}{k},\beta^\vee+\frac{\mu}{k}\right) +2\pi ik \left( u,\beta^\vee+\frac{\mu}{k}\right)\right)}.\label{gsimpletheta1}
\end{align}
Here $\Lambda^{R}\equiv \sum_{\fr{j}=1}^{\text{rank}(\fr{g})}\mathbb{Z}\alpha^\vee_{\fr{j}}$ is the co-root lattice of $\fr{g}$, where $\alpha^\vee_{\fr{j}}$ are the simple co-roots.  $u\in \Lambda^R$, and $\mu$ is a weight. Since the Cartan matrix for a simple Lie algebra has integer entries, $(\alpha^\vee_{{\fr{j}}},\alpha^\vee_{{\fr{l}}})\in\mathbb{Z}$, the theta function \eqref{gsimpletheta1} has the quasi-periodicity
\begin{align}
\begin{cases}
\theta_{\mu,k}(u+m_{\fr{j}}\alpha^\vee_{\fr{j}},\tau)&=\exp{\left( 2\pi i m_{\fr{j}}( \alpha^\vee_{\fr{j}},\mu) \right)}\theta_{\mu,k}(u,\tau),\;m_{\fr{j}},n_{\fr{j}}\in\mathbb{Z}, \\
\theta_{\mu,k}(u+n_{\fr{j}}\alpha^\vee_{\fr{j}},\tau)&= \exp{\left( -\pi i \tau( n_{\fr{j}}\alpha^\vee_{\fr{j}},n_{\fr{j}}\alpha^\vee_{\fr{j}})-2\pi i (u, n_{\fr{j}}\alpha^\vee_{\fr{j}})\right)}\;\theta_{\mu,k}(u,\tau).
\end{cases}
\end{align}
In particular, if $\mu$ is an integral weight, i.e. $\mu_{\fr{j}}\equiv (\alpha^\vee_{\fr{j}},\mu)\in\mathbb{Z}$, then $\theta_{\mu,k}(u+m_{\fr{j}}\alpha^\vee_{\fr{j}},\tau)=\theta_{\mu,k}(u,\tau)$.

\section{Quadratic Differentials}\label{secquaddiff}
We summarize essential facts about quadratic differentials on a Riemann surface from Strebel \cite{strebel1984} and Hubbard \& Masur \cite{hubbard1979}. 

Consider a compact Riemann surface $\Sigma$ of genus $g$ and $n$ punctures, endowed with a complex structure 
which defines a local complex coordinate denoted by $z$. A (meromorphic) \textbf{quadratic differential} $\varphi$ on $\Sigma$ is a $(2,0)$-meromorphic differential; it locally takes the form
\begin{align}
\varphi=h(z)dz\otimes dz\equiv h(z)dz^2,
\end{align}
where $h(z)$ is meromorphic, and under a holomorphic change of coordinate $z\rightarrow \tilde{z}(z)$, it transforms by the chain rule as
\begin{align}
z\rightarrow \tilde{z}(z),\quad h(z)&\rightarrow\tilde{h}(\tilde{z})=\left( \frac{d z}{d\tilde{z}}\right)^2 h(z),\text{ so that } \varphi=\tilde{h}(\tilde{z}) d\tilde{z}^2=h(z)dz^2.
\end{align}
When $h(z)$ is holomorphic, then $\varphi$ is a \textbf{holomorphic quadratic differential}. On a closed genus-$g>1$ Riemann surface without punctures, the complex dimension of the space of all holomorphic quadratic differentials is $(3g-3)$, as a result of the Riemann-Roch theorem.

%
Quadratic differentials find applications in physics, especially in conformal field theory and string field theory (see e.g. \cite{Giddings:1986rf,Giddings:1987im,Carlip:1989xp,Sonoda:1989wa}), because they provide a convenient foliation for
a Riemann surface $\Sigma$. Given a meromorphic quadratic differential $\varphi$, a \textbf{horizontal trajectory} is a non-self-intersecting continuous loop on which $\varphi$ is real and positive, while a \textbf{vertical trajectory} is a non-self-intersecting continuous loop on which $\varphi$ is real and negative. Equivalently, on a local patch $U$ of $\Sigma$ with complex coordinate $z$, and a base point $p_0\in U$, we can define a local natural complex coordinate $w$ on $p\in U$ by
\begin{align}
w(p)&\equiv \int_{p_0}^p \sqrt{h(z)} dz\quad,\quad \varphi=h(z)dz^2.
\end{align}
Then, on a horizontal (vertical) trajectory, $w$ has  constant imaginary (real) part. A \textbf{critical point} of a 
$\varphi$ meromorphic on $\Sigma$ is a zero or a pole of $\varphi$, while all other points on $\Sigma$ are called regular points. A \textbf{critical trajectory} is a horizontal trajectory that joins critical points. In general, a zero of order $n$ is the endpoint of some $(n+2)$ critical trajectories. 

A quadratic differential $\varphi$ defines a metric on $\Sigma$, which is locally given by
\begin{align}
ds^2&=|h(z)| dzd\bar{z}=\sqrt{h(z)}\sqrt{h^*(\bar{z})}dzd\bar{z},\label{metric}
\end{align}
with the corresponding line element
\begin{align}
|dw|&=\sqrt{|h(z)|}|dz|.
\end{align}
Since $h(z)$ is holomorphic away from critical points, the metric \eqref{metric} is flat away from critical points, while the curvature at a critical point is singular.

A \textit{meromorphic} quadratic differential $\varphi$ on $\Sigma$ is called (Jenkins-)\textbf{Strebel}\footnote{Some parts of the mathematical literature, e.g. Hubbard \& Masur \cite{hubbard1979}, further restrict a Strebel differential to be holomorphic.} if it has ``almost only'' closed horizontal trajectories, i.e. if its non-closed horizontal trajectories cover a set of measure zero. 

There are various existence and uniqueness theorems for quadratic differentials on a Riemann surface, with or without punctures. 
One of them is Theorem 21.1 of \cite{strebel1984} which states

\begin{theorem}
 Consider a closed genus $g>1$ Riemann surface $\Sigma$ without punctures. Let $\{\gamma_i\}$ ($i=1,\ldots,3g-3$) be a system of non-self-intersecting continuous closed loop, which are homotopically non-trivial on $\Sigma$, mutually disjoint and belong to different homotopic classes. Also let $m_{i=1,\ldots,3g-3}>0$. Then there exists a holomorphic Strebel differential $\varphi$ on $\Sigma$ that divides $\Sigma$ into cylinders, each of modulus $M_i=Km_i$, where $K$ is a positive constant independent of $i$. 
 \end{theorem}
 For $g=1$, i.e. the torus, a Strebel differential $\varphi$ obviously exists: $\varphi=dz^2$.

\end{appendix}

\end{document}